\documentstyle[psfig]{mn}

\newcommand{\beq}{\begin{equation}}
\newcommand{\eeq}{\end{equation}}
\newcommand{\DS}{\displaystyle}
\newcommand{\TS}{\textstyle}

\newcommand{\SSS}{\scriptscriptstyle}
\newcommand{\th}{\thinspace}

\newcommand{\OVS}{\mbox{OVS}}
\newcommand{\Chi}{{\raise 0.4ex\hbox{$\chi$}}}
\newcommand{\sub}[1]{_{\raise -0.3ex\hbox{$\scriptstyle #1$}}}
\newcommand{\rmn}[1] {{\rm #1}}

\newcommand{\emphh}[1]{{\em #1}}

\newcommand{\dov}{\mbox{$\delta_{\rm ov}$}}

\newcommand{\LZAMS}{\mbox{$L_\rmn{ZAMS}$}}
\newcommand{\LMS}{\mbox{$L_\rmn{MS}$}}
\newcommand{\LTMS}{\mbox{$L_\rmn{TMS}$}}
\newcommand{\LBAGB}{\mbox{$L_\rmn{BAGB}$}}
\newcommand{\LBGB}{\mbox{$L_\rmn{BGB}$}}
\newcommand{\LHeI}{\mbox{$L_\rmn{HeI}$}}
\newcommand{\LminHe}{\mbox{$L_\rmn{min,He}$}}
\newcommand{\LZAHB}{\mbox{$L_\rmn{ZAHB}$}}
\newcommand{\LZHe}{\mbox{$L_\rmn{ZHe}$}}

\newcommand{\Mc}{\mbox{$M_\rmn{c}$}}
\newcommand{\McBAGB}{\mbox{$M_\rmn{c,BAGB}$}}
\newcommand{\McBGB}{\mbox{$M_\rmn{c,BGB}$}}
\newcommand{\McDU}{\mbox{$M_\rmn{c,{\SSS DU}}$}}
\newcommand{\McHeI}{\mbox{$M_\rmn{c,HeI}$}}
\newcommand{\Mcmax}{\mbox{$M_\rmn{c,max}$}}
\newcommand{\McSN}{\mbox{$M_\rmn{c,SN}$}}
\newcommand{\Mche}{\mbox{$M_\rmn{c,He}$}}
\newcommand{\Mcco}{\mbox{$M_\rmn{c,CO}$}}
\newcommand{\MCh}{\mbox{$M_{\rm Ch}$}}

\newcommand{\Mec}{\mbox{$M_{\rm ec}$}}
\newcommand{\MFGB}{\mbox{$M_{\rm FGB}$}}
\newcommand{\Mhook}{\mbox{$M_{\rm hook}$}}

\newcommand{\MHeF}{\mbox{$M_{\rm HeF}$}}
\newcommand{\MWD}{\mbox{$M_{\rm WD}$}}
\newcommand{\MNS}{\mbox{$M_{\rm NS}$}}
\newcommand{\Mup}{\mbox{$M_{\rm up}$}}
\newcommand{\Menv}{\mbox{$M_{\rm env}$}}
\newcommand{\RZAMS}{\mbox{$R_\rmn{ZAMS}$}}
\newcommand{\RMS}{\mbox{$R_\rmn{MS}$}}
\newcommand{\RTMS}{\mbox{$R_\rmn{TMS}$}}
\newcommand{\Ragb}{\mbox{$R_\rmn{AGB}$}}

\newcommand{\RBH}{\mbox{$R_\rmn{BH}$}}
\newcommand{\Rgb}{\mbox{$R_\rmn{GB}$}}
\newcommand{\RHeI}{\mbox{$R_\rmn{HeI}$}}
\newcommand{\RmHe}{\mbox{$R_\rmn{mHe}$}}
\newcommand{\Rmin}{\mbox{$R_\rmn{min}$}}
\newcommand{\RNS}{\mbox{$R_\rmn{NS}$}}
\newcommand{\RWD}{\mbox{$R_\rmn{WD}$}}
\newcommand{\RZAHB}{\mbox{$R_\rmn{ZAHB}$}}
\newcommand{\RZHe}{\mbox{$R_\rmn{ZHe}$}}
\newcommand{\tbl}{\mbox{$\tau_\rmn{bl}$}}
\newcommand{\tMS}{\mbox{$t_\rmn{MS}$}}
\newcommand{\tBAGB}{\mbox{$t_\rmn{BAGB}$}}
\newcommand{\tBGB}{\mbox{$t_\rmn{BGB}$}}
\newcommand{\tHe}{\mbox{$t_\rmn{He}$}}
\newcommand{\tHeI}{\mbox{$t_\rmn{HeI}$}}
\newcommand{\tHeMS}{\mbox{$t_\rmn{HeMS}$}}

\newcommand{\Teff}{\mbox{$T_\rmn{eff}$}}

\newcommand{\Msun}{\hbox{$\th {\rm M}_{\odot}$}}
\newcommand{\Lsun}{\hbox{$\th {\rm L}_{\odot}$}}

\title{Comprehensive analytic formulae for stellar evolution as a function 
of mass and metallicity} 

\author[J. R. Hurley, O. R. Pols and C. A. Tout]
  {Jarrod R. Hurley$^1$,
  Onno R. Pols$^2$
  and Christopher A. Tout$^1$\\
  $^1$Institute of Astronomy, Madingley Road, Cambridge CB3 0HA, UK \\
  $^2$Instituto de Astrof\'{\i}sica de Canarias, c/ Via L\'actea s/n,
  E-38200 La Laguna, Tenerife, Spain \\
  E-mail: {\rm jhurley@ast.cam.ac.uk, onno@ll.iac.es, cat@ast.cam.ac.uk} }

\pagerange{\pageref{firstpage}--\pageref{lastpage}}
\pubyear{1999}

\begin{document}
\label{firstpage}

\maketitle

\begin{abstract}

We present analytic formulae that approximate the evolution of stars for
a wide range of mass $M$ and metallicity $Z$. Stellar luminosity, radius
and core mass are given as a function of age, $M$ and $Z$, for all phases 
from the zero-age main-sequence up to, and including, the remnant stages. 
For the most part we find continuous formulae accurate to within 5\% of 
detailed models. 
These formulae are useful for purposes such as population synthesis that 
require very rapid but accurate evaluation of stellar properties, 
and in particular for use in combination with $N$-body codes.
We describe a mass loss prescription that can be used with these formulae 
and investigate the resulting stellar remnant distribution. 

\end{abstract}


\section{Introduction}
\label{s:intro}

The results of detailed stellar evolution calculations are required for
applications in many areas of astrophysics.
Examples include modelling the chemical evolution of galaxies, determining
the ages of star clusters and simulating the outcomes of stellar collisions.
As stellar evolution theory, and our ability to model it, is continually
being improved (the treatment of convective overshooting and
thermal pulses, for example) there is an ongoing need to update the results
of these calculations.
For a recent overview of problems in stellar evolution see
Noels et al.\ (1995).

As with all theories our understanding of stellar evolution must be
tested against observations.
One way to do this is to attempt to reproduce the findings of large-scale
star surveys, such as the Bright Star Catalogue (Hoffleit 1983) and the
Hipparcos Catalogue (Perryman et al.\ 1997), using population synthesis.
The Hipparcos Catalogue is an excellent example of how improved observing
techniques can initiate a re-evaluation of many aspects of stellar evolution
theory (Baglin 1997; de Boer et al.\ 1997; Van Eck et al.\ 1998).
In order to make population synthesis statistically meaningful it is
necessary to evolve a large sample of stars so as to overcome Poisson noise. 
If we synthesize $n$ examples of a particular type of star we have an 
error of $\pm \sqrt{n}$ which means that for rarer stars often millions of 
possible progenitors are required to get a sufficently accurate sample. 
However, detailed evolution codes can take several hours to evolve a model
of just one star.
Thus it is desirable to generate a large set of detailed models and present
them in some convenient form in which it is relatively simple to utilise
the results at a later stage.

There are two alternative approaches to the problem of using the output of a
series of stellar-evolution runs as data for projects that require them.
One approach is to construct tables (necessarily
rather large, especially if a range of metallicities and/or overshoot parameter
is to be incorporated) and interpolate within these tables.
The other is to approximate the data by a number of interpolation formulae
as functions of age, mass and metallicity.
Both procedures have advantages and disadvantages (Eggleton 1996),
so we have worked on both simultaneously.
Stellar models have been available in tabular form for many years 
(Schaller et al.\ 1992; Charbonnel et al.\ 1993; Mowlavi et al.\ 1998). 
Stellar populations cover a wide range of metallicity
so the ideal is to have a set of models that cover the full range of possible
compositions and stellar masses.
In a previous paper (Pols et al.\ 1998) we presented the results of stellar
evolution calculations for a wide range of mass and metallicity
in tabular form.
In the present paper we report on the results of the second approach,
construction of a set of single star evolution (SSE) formulae,
thus expanding the work of
Eggleton, Fitchett \& Tout (1989) along the lines of Tout et al.\ (1996).
It is more
difficult in practice to find analytic approximations of a conveniently simple
nature for the highly non-uniform movement of a star in the
Hertzsprung-Russell diagram (HRD) than it
is to interpolate in tables, but the resulting code is very much more
compact and adaptable to the requirements of, for example, an $N$-body code
(Aarseth 1996) or variable mass loss.
This is reinforced in the circumstance where one wishes to
include binary-star interactions, such as Roche-lobe overflow, common-envelope
evolution, and magnetic braking with tidal friction, for example (Tout et al.\
1997). 

In Section~\ref{s:sevovw} we provide a brief overview of how stars behave as 
they evolve in time which introduces some of the terminology that we use 
and will hopefully facilitate the understanding of this paper. 
Section~\ref{s:models} describes the detailed models from which the formulae 
are derived and justifies the inclusion of enhanced mixing processes. 
In Section~\ref{s:proc} we outline the procedure to be used for generating 
the SSE package. 
The evolution formulae are presented in Section~\ref{s:fitfor} for all 
nuclear burning phases from the main-sequence to the asymptotic giant branch. 
Our formulae are a vast improvement on the work of Eggleton, Fitchett \& Tout 
(1989) not only due to the inclusion of metallicity as a free 
parameter but also because we have taken a great deal of effort to 
provide a more detailed and accurate treatment of all phases of the evolution. 
Features such as main-sequence formulae that are continuous over the
entire mass range and the modelling of second dredge-up and thermal pulses 
will be discussed. 
Section~\ref{s:finstg} discusses the behaviour of a star as the stellar 
envelope becomes small in mass and outlines what happens when the nuclear 
evolution is terminated.  
We also provide formulae which model the subsequent remnant 
phases of evolution. 
In Section~\ref{s:mlossrt} we describe a comprehensive mass loss algorithm 
which can be used in conjunction with the evolution formulae, as well as 
a method for modelling stellar rotation. 
Various uses for the formulae and future improvements are discussed in 
Section~\ref{s:disc} along with details of how to obtain the formulae in  
convenient subroutine form. 

\section{Stellar evolution overview}
\label{s:sevovw}

A fundamental tool in understanding stellar evolution is the
Hertzsprung-Russell diagram (HRD) which provides a
correlation between the observable stellar properties of luminosity, $L$,
and effective surface temperature, \Teff.
Figure~\ref{f:popi} shows the evolution of a selection of stars in the HRD 
from the zero-age main-sequence (ZAMS), where a star adjusts itself to nuclear 
burning equilibrium, until the end of their nuclear burning lifetimes. 
As stars take a relatively short time to reach the ZAMS 
all ages are measured from this point. 
The length of a stars life, its path on the HRD and its ultimate fate
depend critically on its mass. 

Stars spend most of their time on or near the main-sequence (MS) 
burning hydrogen to produce helium in their cores. 
To first order, the behaviour of a star on the MS can be linked to whether
it has a radiative or convective core. 
Stars with $M \la 1.1 \Msun$ have radiative cores while in higher mass 
stars a convective core develops as a result of the steep temperature 
gradient in the interior. 
During core hydrogen burning on the
MS, low-mass stars will move upwards in $L$ and to higher \Teff\ on the HRD
while higher mass stars will also move upwards in $L$ but to a region of
lower \Teff.
The MS evolution will end when the star has exhausted its supply of hydrogen
in the core.
Low-mass stars will continue expanding as they evolve off the MS but 
for higher mass stars with convective cores the transition is not so smooth.
Owing to mixing in the core there is a sudden depletion of fuel over a large
region which leads to a rapid contraction over the inner region at
core hydrogen-exhaustion.
This causes the hydrogen-exhausted phase gap, or MS hook, which occurs on a
thermal timescale.
The different features of MS evolution are illustrated
by comparing the evolution tracks for the $1.0 \Msun$ and $1.6 \Msun$ stars
in Figure~\ref{f:popi}.

The immediate post-MS evolution towards the right in the HRD occurs at nearly 
constant luminosity and is very rapid. 
For this reason very few stars are seen in this phase, and this region of the
HRD is called the Hertzsprung gap (HG), or the sub-giant branch.
During this HG phase the radius of the star increases greatly causing a
decrease in \Teff.
For cool envelope temperatures the opacity increases causing a
convective envelope to develop.
As the convective envelope grows in extent the star will reach the giant
branch (GB) which is the nearly vertical line corresponding to a fully
convective star, also known as the Hayashi track.
All stars ascend the GB with the hydrogen-exhausted core contracting slowly in
radius and heating while the hydrogen-burning shell is eating its way outwards
in mass and leaving behind helium to add to the growing core.
As the stars move up the GB convection extends over an increasing portion 
of the star. 
The convective envelope may even reach into the previously burnt (or 
processed) regions so that burning products are mixed to the surface in a 
process called dredge-up.

Eventually a point is reached on the GB where the core temperature is
high enough for stars to ignite their central helium supply.
For massive stars, $M \ga 2.0 \Msun$, this takes place gently.
When core helium burning (CHeB) begins the 
star descends along the GB until contraction
moves the star away from the fully convective region of the HRD and back
towards the MS in what is called a blue loop.
During CHeB, carbon and oxygen are produced in the core.
Eventually core helium is exhausted and the star moves back to the right in
the HRD.
The size of the blue loop generally increase with mass, as can
be seen by comparing the $4.0 \Msun$ and $10.0 \Msun$ tracks in 
Figure~\ref{f:popi}.
Lower mass stars have degenerate helium cores on the GB leading to an abrupt 
core-helium flash at helium ignition (HeI).
The star then moves down to the zero-age horizontal branch (ZAHB)
very quickly.
The initial position of a star along the ZAHB depends on the mass of the
hydrogen-exhausted core at the time of ignition and on the mass 
in the overlying envelope.
Those stars with lower mass, ie. shallower envelopes, appear bluer because 
there is less mass to shield the hot hydrogen burning shell.
It is also possible for stars of very high mass, $M \ga 12.0 \Msun$,
to reach high enough central temperatures on the HG for helium to ignite 
before reaching the GB. 
The $16.0 \Msun$ star in Figure~\ref{f:popi} is such an example.
As a result these stars by-pass the GB phase of evolution.

Evolution after the exhaustion of core-helium is very similar to 
evolution after core-hydrogen exhaustion at the end of the MS.
The convective envelope deepens again so that the star once more moves
across towards the Hayashi track to begin what is called the asymptotic
giant branch (AGB).
On the AGB the star consists of a dense core composed of carbon and oxygen
surrounded by a helium burning shell which adds carbon to the
degenerate core.
Initially the H-burning shell is extinguished so that the luminosity is 
supplied exclusively by the He-burning shell; characterizing the early 
AGB (EAGB) phase. 
If the star is massive enough the convective envelope can reach into the 
H-exhausted region again (second dredge-up). 
When the He-burning shell catches up with the H-rich envelope the H-shell 
reignites and the two grow together with the H-burning shell supplying most 
of the luminosity. 
During the following phase the helium shell is unstable,  
which can cause a helium shell flash in which the helium shell will
suddenly release a large amount of luminosity.
The energy released in the flash expands the star resulting in the hydrogen
shell cooling so much that it is extinguished.
Convection once again reaches downward past the dead hydrogen shell. 
This mixes helium to the surface, as well as carbon that was mixed out of the
helium shell by flash-driven convection.
As the star subsequently contracts the convection recedes and the hydrogen 
shell re-ignites but has now moved inwards in mass due to the envelope
convection.
This process is called third dredge-up. 
The star continues its evolution up the AGB with the hydrogen shell
producing almost all of the luminosity.
The helium shell flash can repeat itself many times and the cycle is known 
as a thermal pulse. 
This is the thermally pulsing asymptotic giant branch (TPAGB).  

The stellar radius can grow to very large values on the AGB which lowers
the surface gravity of the star, so that the surface material is less
tightly bound.
Thus mass loss from the stellar surface can become significant with the
rate of mass loss actually accelerating with time during continued
evolution up the AGB.
Unfortunately, our understanding of the mechanisms that cause this mass
loss is poor with possible suggestions linking it to the helium shell flashes
or to periodic envelope pulsations.
Whatever the cause, the influence on the evolution of AGB stars
is significant.
Mass loss will eventually remove all of the stars envelope so that the
hydrogen burning shell shines through.
The star then leaves the AGB and evolves to hotter \Teff\ at nearly
constant luminosity.
As the photosphere gets hotter the energetic photons become absorbed by the
material which was thrown off while on the AGB.
This causes the material to radiate and the star may be seen as a planetary
nebula.
The core of the star then begins to fade as the nuclear burning ceases. 
The star is now a white dwarf (WD) and cools slowly at high temperature
but low luminosity.

If the mass of the star is large enough, $M \ga 7 \Msun$,
the carbon-oxygen core is not degenerate and will ignite carbon as it 
contracts, followed by a succession of nuclear reaction sequences which very 
quickly produce an inner iron core. 
Any further reactions are endothermic and cannot contribute to the 
luminosity of the star. 
Photodisintegration of iron, combined with electron capture by protons and 
heavy nuclei, then removes most of the electron degeneracy pressure that was 
supporting the core and it begins to collapse rapidly. 
When the density becomes large enough the inner core rebounds 
sending a shockwave outwards through the outer layers 
of the star that have remained suspended above the collapsing core. 
As a result the envelope of the star is ejected in a supernova (SN) 
explosion so that the AGB is effectively truncated at the start of carbon  
burning and the star has no TPAGB phase.  
The remnant in the inner core will stablise to form a neutron star (NS) 
supported by neutron degeneracy pressure unless the initial stellar mass 
is large enough that complete collapse to a black hole (BH) occurs. 

Stars with $M \ga 15 \Msun$ are severely affected by mass loss during their 
entire evolution and may lose their envelopes during CHeB, or even on the 
HG, exposing nuclear processed material. 
If this occurs then a naked helium star is produced and such stars, or stars 
about to become naked helium stars, may be Wolf-Rayet stars.  
Wolf-Rayet stars are massive objects which are found near the MS, are losing 
mass at very high rates and show weak, or no, hydrogen lines in their spectra. 
Luminous Blue Variables (LBVs) are extremely massive post-MS objects 
with enormous mass loss rates in a stage of evolution just prior to becoming 
a Wolf-Rayet star. 
Naked helium stars can also be produced from less massive stars in binaries 
as a consequence of mass transfer. 

Variations in composition can also affect the stellar evolution timescales as 
well as the appearance of the evolution on the HRD, and even the ultimate fate 
of the star.
A more detailed discussion of the various phases of evolution can be found 
throughout this paper.

\section{Stellar models}
\label{s:models}

The fitting formulae are based on the stellar models computed by
Pols et al.\ (1998).
They computed a grid of evolution tracks for masses $M$ between 0.5 and
50\Msun\ and for seven values of metallicity,
$Z=0.0001$, 0.0003, 0.001, 0.004, 0.01, 0.02 and 0.03.
They also considered the problem  of enhanced mixing such as overshooting 
beyond the classical boundary of convective instability.
Its effect was modelled with a prescription based on a modification of the 
Schwarzschild stability criterion, introducing a free parameter \dov\
(which differs from the more commonly used parameter relating the
overshooting distance to the pressure scale height; see Pols et al.\ 1998 
for details).
The tracks computed with a moderate amount of enhanced mixing (given by 
$\dov=0.12$ and labeled the OVS tracks by Pols et al.\ 1998) were
found to best reproduce observations in a series of sensitive tests
involving open clusters and ecliping binaries (see Schr\"oder, Pols \& 
Eggleton 1997; Pols et al.\ 1997, 1998). 
We consequently use these OVS tracks as the data to which we fit our formulae. 

For each $Z$, 25 tracks were computed spaced by approximately 0.1 in $\log M$,
except between 0.8 and 2.0\Msun where four extra models were added 
to resolve the shape of the main-sequence which changes rapidly 
in this mass range.
Hence we dispose of a database of 175 evolution tracks, each containing
several thousand individual models.

\begin{figure}
\centerline{
\psfig{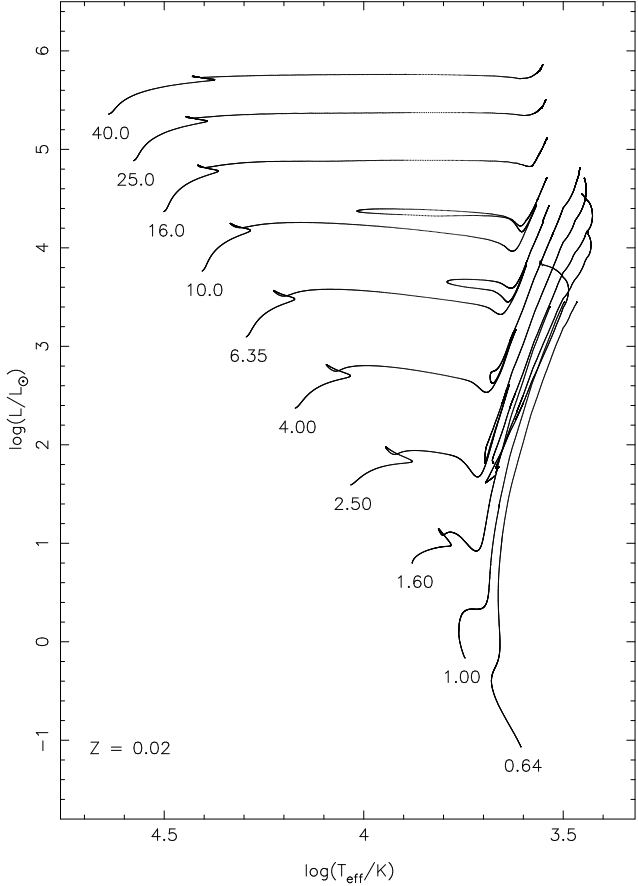}
}
\caption{Selected \OVS\ evolution tracks for $Z = 0.02$,
for masses 0.64, 1.0, 1.6, 2.5, 4.0, 6.35, 10, 16, 25 and 40\Msun.}
\label{f:popi}
\end{figure}

\begin{figure}
\centerline{
\psfig{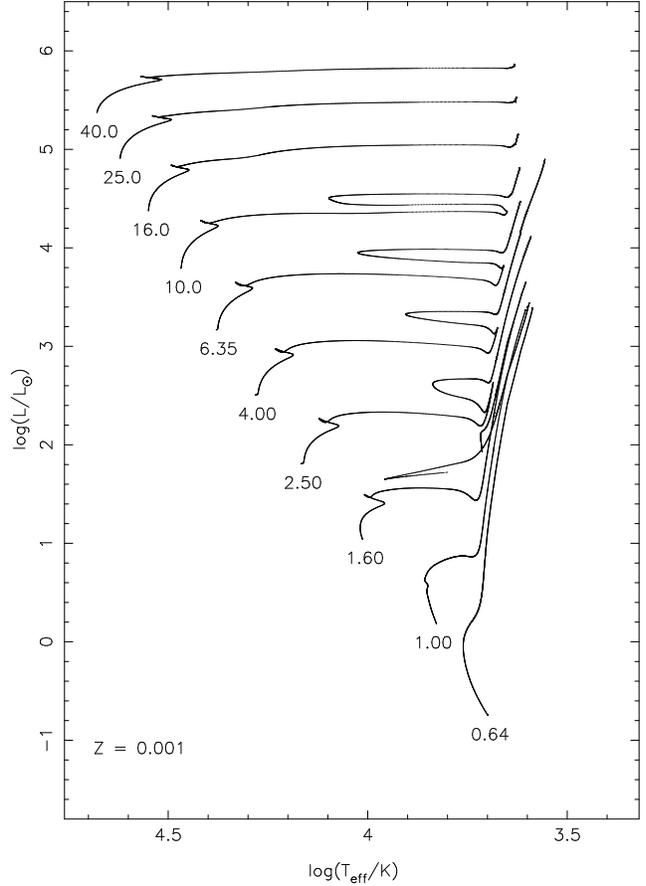}
}
\caption{Same as Fig.~\ref{f:popi} for $Z = 0.001$. The 1.0\Msun\ post
He flash track has been omitted for clarity.}
\label{f:popii}
\end{figure}

\begin{figure}
\centerline{
\psfig{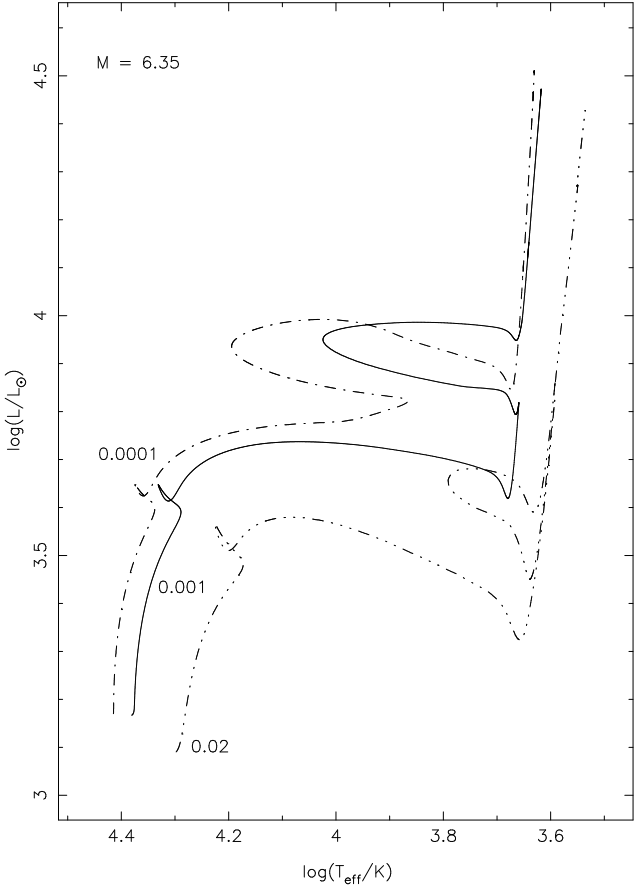}
}
\caption{Detailed \OVS\ evolution tracks for $M = 6.35 \Msun$,
for metallicities 0.0001, 0.001 and 0.02.}
\label{f:ffig3}
\end{figure}

\begin{figure}
\centerline{
\psfig{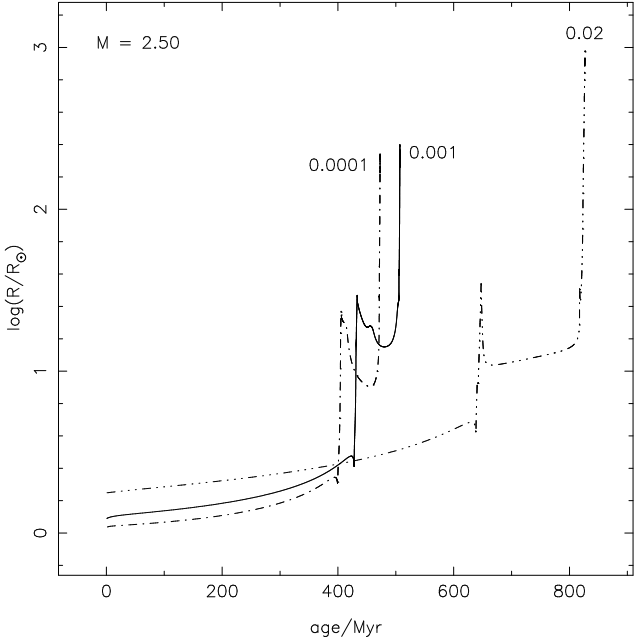}
}
\caption{Radius evolution as a function of stellar age for $M = 2.5 \Msun$,
for metallicities 0.0001, 0.001 and 0.02. Tracks are from the detailed
models and run from the ZAMS to the point of termination on the AGB.}
\label{f:ffig4}
\end{figure}

A subset of the resulting \OVS\ tracks in the HRD are shown in
Fig.~\ref{f:popi} for $Z=0.02$ and Fig~\ref{f:popii} for $Z=0.001$. 
The considerable variation of model behaviour introduced by changes in
metallicity is illustrated by Fig.~\ref{f:ffig3}.
Detailed models of the same mass, $M = 6.35 \Msun$, are shown on the HRD for
three different metallicities, $Z = 0.0001, 0.001$ and 0.02.
Not only does a change in composition move the track to a different position
in the HRD but it also changes the appearance of each track, as can be seen
by considering the extent of the hook feature towards the end of the main 
sequence and the blue loops during core helium burning. 
Furthermore, the $Z = 0.0001$ model ignites helium in its core while on the
Hertzsprung gap as opposed to the other models which evolve up the giant
branch before reaching a high enough core temperature to start helium
burning.
In addition the nuclear burning lifetime of a star can change by as much as
a factor of 2 owing to differences in composition, as shown in
Fig.~\ref{f:ffig4} for a set of $2.5 \Msun$ models.
This emphasizes the need to present the results of stellar evolution
calculations for an extensive range of metallicity.

Mass loss from stellar winds was neglected in the detailed stellar models,
mainly because the mass loss rates are uncertain by at least a factor of 
three.  We do include mass loss in our analytic formulae in an elegant way, 
as will be described in Section~\ref{s:mloss}, which allows us to experiment
easily with different mass loss rates and prescriptions.  

\section{Procedure}
\label{s:proc}

We assign each evolution phase an 
integer type, $k$, where:
\begin{eqnarray*}
0 & = & \mbox{ MS star } M \la 0.7 \mbox{ deeply or fully convective} \\
1 & = & \mbox{ MS star } M \ga 0.7 \\
2 & = & \mbox{ Hertzsprung Gap (HG)} \\
3 & = & \mbox{ First Giant Branch (GB)} \\
4 & = & \mbox{ Core Helium Burning (CHeB)} \\
5 & = & \mbox{ Early Asymptotic Giant Branch (EAGB)} \\
6 & = & \mbox{ Thermally Pulsing Asymptotic Giant Branch (TPAGB)} \\
7 & = & \mbox{ Naked Helium Star MS (HeMS)} \\
8 & = & \mbox{ Naked Helium Star Hertzsprung Gap (HeHG)} \\
9 & = & \mbox{ Naked Helium Star Giant Branch (HeGB)} \\
10 & = & \mbox{ Helium White Dwarf (He WD)} \\
11 & = & \mbox{ Carbon/Oxygen White Dwarf (CO WD)} \\
12 & = & \mbox{ Oxygen/Neon White Dwarf (ONe WD)} \\
13 & = & \mbox{ Neutron Star (NS)} \\
14 & = & \mbox{ Black Hole (BH)} \\
15 & = & \mbox{ massless remnant,}
\end{eqnarray*}
and we divide the MS into two phases to distinguish between deeply or fully 
convective low-mass stars and stars of higher mass with little or no convective 
envelope as these will respond differently to mass loss. 

To begin with we take different features of the evolution in turn,
e.g. MS lifetime, ZAHB luminosity,
and first try to fit them as $f \left( M \right)$ for a particular $Z$ in
order to get an idea of the functional form.
We then extend the function to $g \left( M, Z \right)$ using
$f \left( M \right)$ as a starting point.
In this way we fit formulae to the end-points of the various evolutionary
phases as well as to the timescales.
We then fit the behaviour within each phase as $h  \left( t, M, Z \right)$,
e.g. $\LMS \left( t, M, Z \right)$.

As a starting point we take the work of Tout et al.~(1996) who fitted the
zero-age main-sequence luminosity ($\LZAMS$) and radius ($\RZAMS$)
as a function of $M$ and $Z$.
Their aim, as is ours, was to find simple computationally efficent functions
which are accurate, continuous and differentiable in $M$ and $Z$,
such as rational polynomials.
This is acheived using least-squares fitting to the data after choosing
the initial functional form.
In most cases we determine the type of function, the value of the powers, 
and the number of coefficients to be used, simply by inspecting the shape 
of the data, however in some cases, such as the luminosity-core-mass relation 
on the giant branch, the choice will be dictated by an underlying physical 
process. 
For the ZAMS, accuracy is very important because it fixes the star's
position in the HRD.
Tout et al.~(1996) acheived \LZAMS\ accurate to 3\% and \RZAMS\ accurate to
1.2\% over the entire range.
For the remainder of the functions we aim for RMS errors less than 5\%
and preferably a maximum individual error less than 5\% although this has 
to be relaxed for some later stages of the evolution where the behaviour
varies greatly with $Z$ but also where the model points are more uncertain
owing to shortcomings in stellar evolution theory.

\section{Fitting formulae}
\label{s:fitfor}

In this section we present our formulae describing the evolution as
a function of mass $M$ and age $t$. The explicit $Z$-dependence is in most
cases not given here because it would clutter up the presentation.  This
$Z$-dependence is implicit whenever a coefficient of the form $a_n$ or
$b_n$ appears in any of the formulae.
The explicit dependence of these coefficients on $Z$
is given in the Appendix. Coefficients of the form $c_n$, whose numerical
values are given in this section, do not depend on $Z$.

We adopt the following unit conventions: numerical values of
mass, luminosity and radius are in solar units, and values of timescales
and ages are in units of $10^6$\,yr, unless otherwise specified.

We begin by giving formulae for the most important critical masses, \Mhook\
(the initial mass above which a hook appears in the main-sequence), \MHeF\
(the maximum initial mass for which He ignites degenerately in a helium
flash) and \MFGB\ (the maximum initial mass for which He ignites on the
first giant branch). Values for these masses are given in Table~1 of Pols et
al.~(1998) estimated from the detailed models for 7 metallicities. These
values can be accurately fitted as a function of $Z$ by the following
formulae, where $\zeta = \log(Z/0.02)$:
\beq
\Mhook = 1.0185 + 0.16015\zeta + 0.0892\zeta^2,
\eeq
\beq
\MHeF = 1.995 + 0.25\zeta + 0.087\zeta^2,
\eeq
\beq
\MFGB = \frac{13.048 \left( Z/0.02 \right)^{0.06}}
{1 + 0.0012 \left( 0.02/Z \right)^{1.27}}.
\eeq
Based on the last two critical masses, we make a distinction into
three mass intervals, which will be useful in the later descriptions:
\begin{enumerate}
\item low-mass (LM) stars, with $M < \MHeF$, develope degenerate He cores
  on the GB and ignite He in a degenerate flash at the top of the GB;
\item intermediate-mass (IM) stars, with $\MHeF \le M < \MFGB$, which evolve
  to the GB without developing degenerate He cores, also igniting He at the
  top of the GB;
\item high-mass (HM) stars, with $M > \MFGB$, ignite He in the HG before the
  GB is reached, and consequently do not have a GB phase.
\end{enumerate}
Note that this definition of IM and HM stars is different from the more often
used one, based on whether or not carbon ignites non-degenerately.

\subsection{Main-sequence and Hertzsprung gap}
\label{s:MSHG}

To determine the base of the giant branch (BGB) we find where the
mass of the convective envelope $M_{\rm CE}$ first exceeds a set fraction
of the envelope mass $M_{\rm E}$ as $M_{\rm CE}$ increases on the HG.
From inspection the following fractions
\begin{eqnarray*}
M_{\rm CE} & = & \frac{2}{5} M_{\rm E} \quad M \la \MHeF \\
M_{\rm CE} & = & \frac{1}{3} M_{\rm E} \quad M \ga \MHeF
\end{eqnarray*}
generally give a BGB point corresponding to the local minimum in
luminosity at the start of the GB.
We define helium ignition as the point where $L_\rmn{He} = 0.01 L$ for the
first time.
For HM stars this will occur before the BGB point is found, ie. no GB,
and thus we set $\tBGB = \tHeI$ for the sake of defining an end-point to
the HG, so that BGB is more correctly the end of the HG (EHG) as this is true 
over the entire mass range. 

\begin{figure}
\centerline{
\psfig{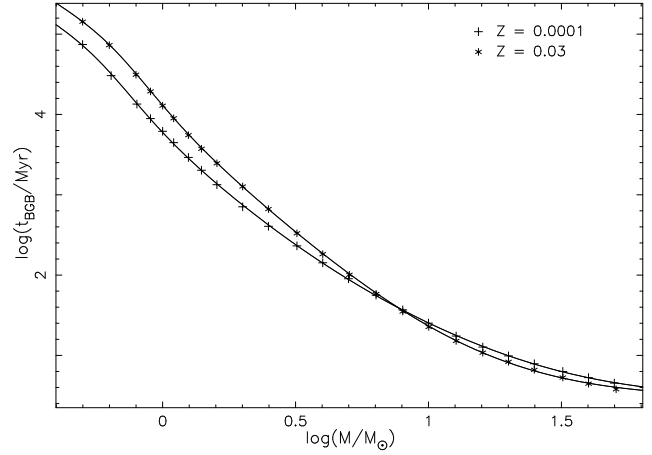}
}
\caption{Time taken to reach the base of the giant branch as a function
of stellar mass as given by eq.~(\ref{e:tBGB}), shown against the
detailed model points, for $Z = 0.0001$ and 0.03 which give the worst 
fit of all the metallicities. The maximum error 
over the entire metallicity range is 4.8\% and the RMS error is 1.9\%.}
\label{f:ffig5}
\end{figure}

The resultant lifetimes to the BGB are fitted as a function of $M$ and $Z$ by 
\beq\label{e:tBGB}
\tBGB = \frac{a_{\SSS 1} + a_{\SSS 2} M^4 + a_{\SSS 3} M^{5.5} + M^7}
{a_{\SSS 4} M^2 + a_{\SSS 5} M^7} .
\eeq
Figure~\ref{f:ffig5} shows how eq.~(\ref{e:tBGB}) fits the detailed model 
points for $Z = 0.0001$ and 0.03 which are the metallicities which 
lead to the largest errors.  
Over the entire metallicity range the function gives a rms error of 1.9\% 
and a maximum error of 4.8\%. 
In order that the time spent on the HG will always be a small fraction of the 
time taken to reach the BGB, even for low-mass stars which don't have a well 
defined HG, the MS lifetimes are taken to be  
\beq
\tMS = \max \left( t_\rmn{hook}, x \tBGB \right) ,
\eeq
where $t_\rmn{hook} = \mu \tBGB$ and 
\beq
x = \max \left( 0.95, \min \left( 0.95 - 0.03 \left( 
\zeta + 0.30103 \right), 0.99 \right) \right)
\eeq
\beq
\mu = \max \left( 0.5, 1.0 - 0.01 \, \max \left( \frac{a_{\SSS 6}}
{M^{{\TS a}_7}}, 
a_{\SSS 8} + \frac{a_{\SSS 9}}{M^{{\TS a}_{10}}} \right) \right) .
\eeq
Note that $\mu$ is ineffective for $M < \Mhook$, ie. stars without a hook 
feature, and in this case the functions ensure that $x > \mu$.

So we now have defined the time at the end of the MS, \tMS\, and the time
taken to reach the start of the GB (or end of the HG), \tBGB\, such that
\begin{eqnarray*}
t & : & 0.0 \, \rightarrow \, \tMS \qquad \mbox{ MS evolution} \\ 
t & : & \tMS \rightarrow \tBGB \quad \,\, \mbox{ HG evolution.} 
\end{eqnarray*}

The starting values for $L$ and $R$ are the ZAMS points fitted by 
Tout et al.~(1996).
We fit the values at the end of the MS, \LTMS\ and $\RTMS$, as well as at
the end of the HG,
\[ L_\rmn{EHG} = \left\{ \begin{array} {r@{\quad}l}
\LBGB & M < \MFGB \\
\LHeI & M \geq \MFGB
\end{array} \right. \]
\[ R_\rmn{EHG} = \left\{ \begin{array} {r@{\quad}l}
\Rgb \left( \LBGB \right) & M < \MFGB \\
\RHeI & M \geq \MFGB
\end{array} \right. \, . \]

The luminosity at the end of the MS is approximated by 
\beq\label{e:LTMS}
\LTMS = \frac{a_{\SSS 11} M^3 + a_{\SSS 12} M^4 + a_{\SSS 13} 
M^{{\TS a}_{16} + 1.8}}
{a_{\SSS 14} + a_{\SSS 15} M^5 +  M^{{\TS a}_{16}}} 
\eeq
with $a_{\SSS 16} \approx 7.2$.
This proved fairly straightforward to fit but the behaviour of \RTMS\ is 
not so smooth and thus requires a more complicated function in order to 
fit it continuously.
The resulting fit is 
\begin{eqnarray}
\RTMS & = & \frac{a_{\SSS 18} + a_{\SSS 19} M^{{\TS a}_{21}}}
{a_{\SSS 20} + M^{{\TS a}_{22}}} 
\quad M \leq a_{\SSS 17} \label{e:r1tms} \\
\RTMS & = & \frac{c_{\SSS 1} M^3 + a_{\SSS 23} M^{{\TS a}_{26}} + 
a_{\SSS 24} M^{{\TS a}_{26} + 1.5}}
{a_{\SSS 25} + M^5} \, M \geq M_*  , \left( \ref{e:r1tms} \rmn{a} \right) 
\nonumber 
\end{eqnarray}
with straight-line interpolation to connect eqs.~(\ref{e:r1tms}) and 
(\ref{e:r1tms}a) between the endpoints, where 
\[ M_* = a_{\SSS 17} + 0.1 \quad , \quad a_{\SSS 17} \approx 1.4 \] 
and $c_{\SSS 1} = -8.672073 \times 10^{-2}$, $a_{\SSS 21} \approx 1.47$, 
$a_{\SSS 22} \approx 3.07$, $a_{\SSS 26} \approx 5.50$. 
Note that for low masses, $M < 0.5$, where the function is being 
extrapolated we add the condition 
\[ \RTMS = \max \left( \RTMS , 1.5 \RZAMS \right) \] 
to avoid possible trouble in the distant future.

The luminosity at the base of the GB is approximated by 
\beq\label{e:LBGB}
\LBGB = \frac{a_{\SSS 27} M^{{\TS a}_{31}} + a_{\SSS 28} M^{{\TS c}_2}}
{a_{\SSS 29} + a_{\SSS 30} M^{{\TS c}_3} + M^{{\TS a}_{32}}},
\eeq
with $c_{\SSS 2} = 9.301992$, $c_{\SSS 3} = 4.637345$, 
$a_{\SSS 31} \approx 4.60$ and 
$a_{\SSS 32} \approx 6.68$.
The description of $\LHeI$, $\Rgb$ and $\RHeI$ is given in later sections.

\subsubsection{Main-sequence evolution}

On the MS we define a fractional timescale   
\beq
\tau = \frac{t}{\tMS } \, . 
\eeq
As a star evolves across the MS its evolution accelerates so that it's 
possible to model the time dependence of the logarithms of the luminosity and 
radius by polynomials in $\tau$. 
Luminosity is given by 
\begin{eqnarray}\label{e:LMS}
\log \frac{\LMS \left( t \right) }{\LZAMS } & = & 
{\alpha}_{\SSS L} \tau + {\beta}_{\SSS L} {\tau}^{\eta} + 
\left( \log \frac{\LTMS }{\LZAMS } - {\alpha}_{\SSS L} - {\beta}_{\SSS L} 
\right) {\tau}^2 
 \nonumber \\ 
 & & - \Delta L \left( {\tau}_{1}^{2} - {\tau}_{2}^{2} \right)
\end{eqnarray}
and radius by 
\begin{eqnarray}\label{e:RMS}
\lefteqn{ 
\log \frac{\RMS \left( t \right) }{\RZAMS }  =  
{\alpha}_{\SSS R} \tau + {\beta}_{\SSS R} {\tau}^{10} + 
\gamma {\tau}^{40} + 
} \\ 
 & & + \left( \log \frac{\RTMS }{\RZAMS } - {\alpha}_{\SSS R} - 
{\beta}_{\SSS R} - \gamma \right) {\tau}^3 
 - \Delta R \left( {\tau}_{1}^{3} - {\tau}_{2}^{3} \right) \nonumber
\end{eqnarray}
where
\begin{eqnarray}
{\tau}_1 & = & \min \left( 1.0 , t/t_\rmn{hook} \right) \\ 
{\tau}_2 & = & \max \left( 0.0 , \min \left( 1.0 , 
\frac{t - \left( 1.0 - \epsilon \right) t_\rmn{hook}}
{\epsilon \, t_\rmn{hook}} \right) \right) 
\end{eqnarray}
for $\epsilon = 0.01$.

\begin{figure}
\centerline{
\psfig{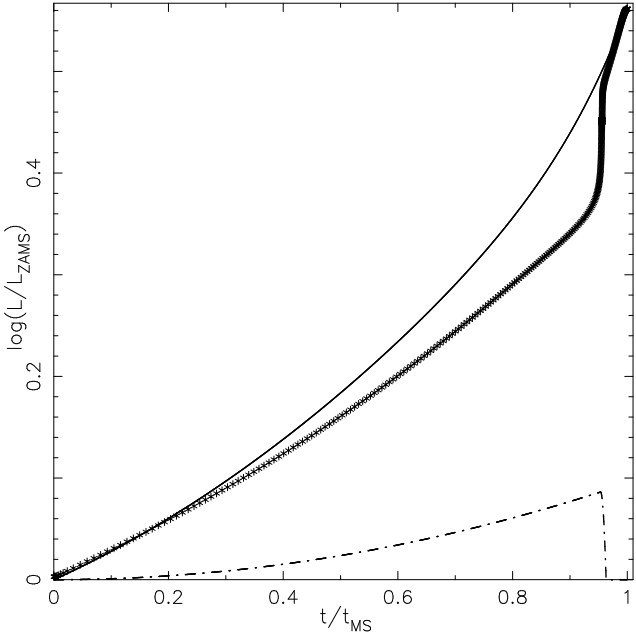}
}
\caption{Luminosity evolution on the main-sequence for a typical detailed
model with a hook feature (star points) decomposed into two functions:
a smooth polynomial (solid line) and a pertubation function (dash-dot line).}
\label{f:ffig6}
\end{figure}

We add $\Delta L$ and $\Delta R$ as pertubations to the smooth polynomial 
evolution of $L$ and $R$ in order to mimic the hook behaviour  
for $M > \Mhook $.
In effect we have 
\[ \LMS \left( t \right) = L_a \left( t \right) / L_b \left( t \right) \] 
where $L_a \left( t \right)$ is a smooth function describing the long-term  
behaviour of $\LMS \left( t \right)$ and $L_b \left( t \right)$ is 
another smooth function describing short-term pertubations where 
\[ \log L_b \left( t \right) = \Delta L \left( {\tau}_{1}^{2} - {\tau}_{2}^{2} 
\right) \] 
and the action of ${\tau}_2$ acheives a smooth transition over 
$ \Delta t = \epsilon \, t_\rmn{hook}$.
This decomposition of $L(t)$ into $L_a(t)$ and $L_b(t)$ for a typical 
detailed model is illustrated by Fig.~\ref{f:ffig6}. 
The luminosity pertubation is approximated by 
\beq
\Delta L = \left\{ \begin{array}{l@{\qquad}l} 
\DS 0.0 & M \leq \Mhook \\ [3ex] 
\DS B \left[ \frac{M - \Mhook }{a_{\SSS 33} - \Mhook } \right]^{0.4} & 
\Mhook < M < a_{\SSS 33} \\ [3ex] 
\DS  \min \left( \frac{a_{\SSS 34}}{M^{{\TS a}_{35}}} , 
\frac{a_{\SSS 36}}{M^{{\TS a}_{37}}} \right) & M \geq a_{\SSS 33} 
\end{array} \right.
\eeq
where $B = \Delta L \left( a_{\SSS 33} \right)$, 
$1.25 < a_{\SSS 33} < 1.4$, $a_{\SSS 35} \approx 0.4$ and 
$a_{\SSS 37} \approx 0.6$.

The radius pertubation is approximated by 
\beq
\Delta R = \left\{ \begin{array}{l@{\:}l} 
\DS 0.0 & M \leq \Mhook \\ [3ex]
\DS a_{\SSS 43} \left( \frac{M - \Mhook }{a_{\SSS 42} - \Mhook } \right)^{0.5} 
& \Mhook < M \leq a_{\SSS 42} \\ 
 & \\ 
\DS a_{\SSS 43} + \left( B - a_{\SSS 43} \right) \left[ 
\frac{M - a_{\SSS 42}}{2.0 - a_{\SSS 42}} 
\right]^{{\TS a}_{44}} & a_{\SSS 42} < M < 2.0 \\ [3ex] 
\DS \frac{a_{\SSS 38} + a_{\SSS 39} M^{3.5}} {a_{\SSS 40} M^3 + 
M^{{\TS a}_{41}}} - 1.0 & M \geq 2.0 
\end{array} \right.
\eeq
where $B = \Delta R \left( M = 2.0 \right)$, 
$a_{\SSS 41} \approx 3.57$, $1.1 < a_{\SSS 42} < 1.25$ and 
$a_{\SSS 44} \approx 1.0$.

The exponent $\eta = 10$ in eq.~(\ref{e:LMS}) unless $Z \leq 0.0009$ when it 
is given by 
\beq
\eta = \left\{ \begin{array} {r@{\quad}l} 
10 & M \leq 1.0 \\ 
20 & M \geq 1.1
\end{array} \right.
\eeq
with linear interpolation between the mass limits. 

\begin{figure}
\centerline{
\psfig{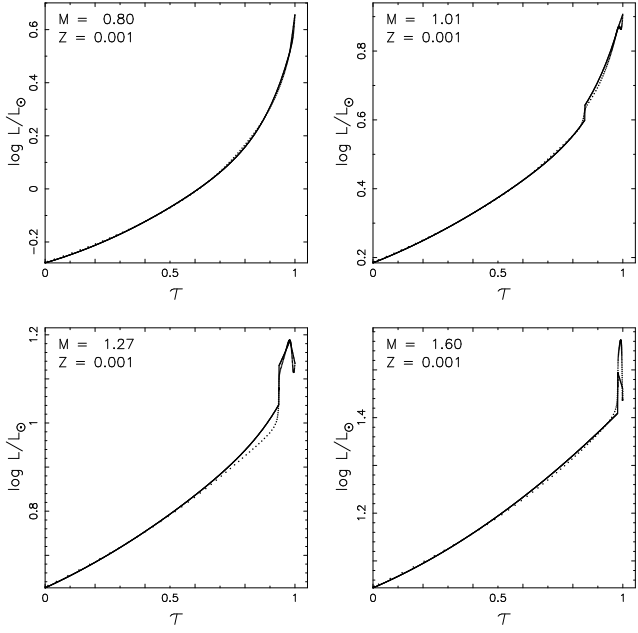}
}
\caption{Luminosity evolution on the main-sequence as given by
eq.~(\ref{e:LMS}) (solid line) and from the detailed models (points)
for selected masses with a metallicity of 0.001.}
\label{f:ffig7}
\end{figure}

\begin{figure}
\centerline{
\psfig{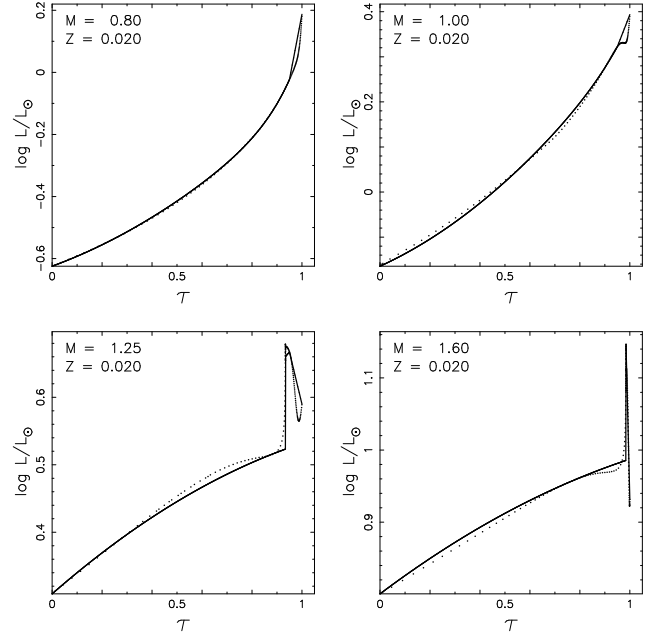}
}
\caption{Same as Fig.~\ref{f:ffig7} for $Z = 0.02$.}
\label{f:ffig8}
\end{figure}

\begin{figure}
\centerline{
\psfig{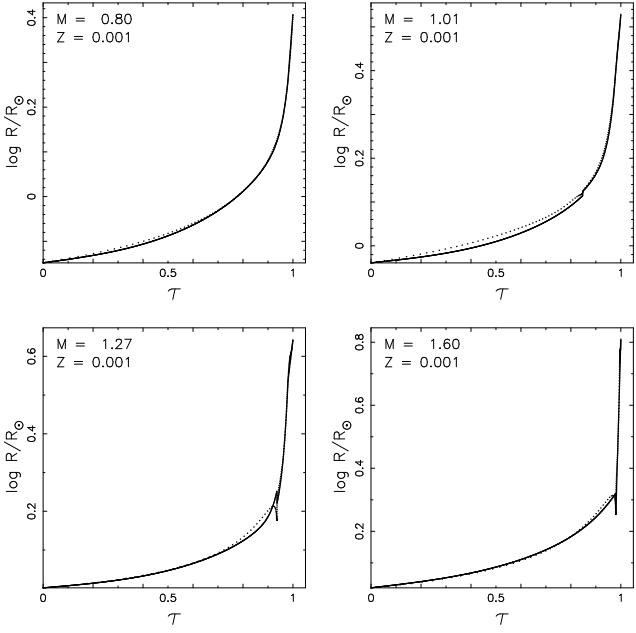}
}
\caption{Radius evolution on the main-sequence as given by
eq.~(\ref{e:RMS}) (solid line) and from the detailed models (points)
for selected masses with a metallicity of 0.001.}
\label{f:ffig9}
\end{figure}

\begin{figure}
\centerline{
\psfig{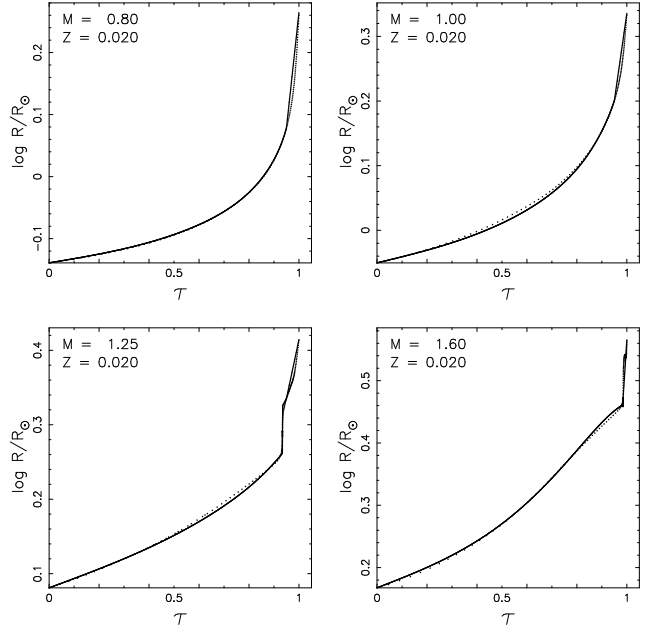}
}
\caption{Same as Fig.~\ref{f:ffig9} for $Z = 0.02$.}
\label{f:ffig10}
\end{figure}

The remaining functions for this section are those that describe the 
behaviour of the coefficients in eqs.~(\ref{e:LMS}) and (\ref{e:RMS}).
The fact that these can appear messy and complicated in places reflects 
rapid changes in the shape of the $L$ and $R$ evolution for the 
detailed models as a function of $M$ as well as $Z$.
This is illustrated in Figs.~\ref{f:ffig7}, \ref{f:ffig8}, \ref{f:ffig9} 
and \ref{f:ffig10} which also show the tracks derived from 
these functions, exhibiting that our efforts  
have not been in vain. 
The fitting of the coefficients is also complicated by the sensitivity of 
eqs.~(\ref{e:LMS}) and (\ref{e:RMS}) to small changes in the values of 
the coefficients.
Ideally we would like all the functions to be smooth and differentiable
across the entire parameter space but in some places this has to be
sacrificed to ensure that the position of all the fitted tracks on the 
HRD is as accurate as possible. 
This is deemed necessary as the main use of the functions is envisaged 
to be the simulation of Colour-Magnitude diagrams for comparison with 
observations.

The luminosity $\alpha$ coefficient is approximated by 
\beq\label{e:Lalpha}
{\alpha}_{\SSS L} = \frac{a_{\SSS 45} + a_{\SSS 46} M^{{\TS a}_{48}}}
{M^{0.4} + a_{\SSS 47} M^{1.9}} 
\qquad M \geq 2.0 \, ,
\eeq
where $a_{\SSS 48} \approx 1.56$, and then 
\[ {\alpha}_{\SSS L} = \left\{ \begin{array} {l@{\qquad\qquad\qquad}l} 
\DS a_{\SSS 49} & \hfill M < 0.5 \\
\multicolumn{2}{l}{
\DS a_{\SSS 49} + 5.0 \left( 0.3 - a_{\SSS 49} \right) \left( M - 0.5 \right) 
} \\ [1ex] 
 & \hfill 0.5 \leq M < 0.7 \\ [2ex]
\multicolumn{2}{l}{
\DS 0.3 + \left( a_{\SSS 50} - 0.3 \right) \left( M - 0.7 \right) /
\left( a_{\SSS 52} - 0.7 \right) 
} \\ [1ex] 
 & \hfill 0.7 \leq M < a_{\SSS 52} \\ [2ex] 
\multicolumn{2}{l}{
\DS a_{\SSS 50} + \left( a_{\SSS 51} - a_{\SSS 50} \right) 
\left( M - a_{\SSS 52} \right) / \left( a_{\SSS 53} - a_{\SSS 52} \right) 
} \\ [1ex] 
 & \hfill a_{\SSS 52} \leq M < a_{\SSS 53} \\ [2ex] 
\multicolumn{2}{l}{
\DS a_{\SSS 51} + \left( B - a_{\SSS 51} \right) 
\left( M - a_{\SSS 53} \right) / \left( 2.0 - a_{\SSS 53} \right) 
} \\ [1ex] 
 & \hfill a_{\SSS 53} \leq M < 2.0 
\end{array} \right. \quad \quad \left( \ref{e:Lalpha} \rmn{a} \right) \]
where $B = {\alpha}_{\SSS L} \left( M = 2.0 \right)$.

The luminosity $\beta$ coefficient is approximated by 
\beq\label{e:Lbeta}
{\beta}_{\SSS L} = \max \left( 0.0 , a_{\SSS 54} - a_{\SSS 55} 
M^{{\TS a}_{56}} \right) 
\eeq
where $a_{\SSS 56} \approx 0.96$. 
Then if $M > a_{\SSS 57}$ and ${\beta}_{\SSS L} > 0.0$ 
\[ {\beta}_{\SSS L} = \max \left( 0.0 , B - 10.0 \left( M - a_{\SSS 57} 
\right) B \right) \] 
where $B = {\beta}_{\SSS L} \left( M = a_{\SSS 57} \right)$ 
and $1.25 < a_{\SSS 57} < 1.4$.

The radius $\alpha$ coefficient is approximated by 
\beq\label{e:Ralpha}
{\alpha}_{\SSS R} = \frac{a_{\SSS 58} M^{{\TS a}_{60}}}
{a_{\SSS 59} M^{{\TS a}_{61}}} 
\qquad a_{\SSS 66} \leq M \leq a_{\SSS 67} \, ,
\eeq
where $a_{\SSS 66} \approx 1.4$ and $a_{\SSS 67} \approx 5.2$, and then 
\[ {\alpha}_{\SSS R} = \left\{ \begin{array} {l@{\qquad}l} 
\DS a_{\SSS 62} & \hfill M < 0.5 \\ [2ex]
\multicolumn{2}{l}{
\DS a_{\SSS 62} + \left( a_{\SSS 63} - a_{\SSS 62} \right) 
\left( M - 0.5 \right) / 0.15 
} \\ [1ex] 
 & \hfill 0.5 \leq M < 0.65 \\ [2ex] 
\multicolumn{2}{l}{
\DS a_{\SSS 63} + \left( a_{\SSS 64} - a_{\SSS 63} \right) 
\left( M - 0.65 \right) / \left( a_{\SSS 68} - 0.65 \right) 
} \\ [1ex] 
 & \hfill 0.65 \leq M < a_{\SSS 68} \\ [2ex] 
\multicolumn{2}{l}{
\DS a_{\SSS 64} + \left( B - a_{\SSS 64} \right) \left( M - a_{\SSS 68} 
\right) / \left( a_{\SSS 66} - a_{\SSS 68} \right) 
} \\ [1ex] 
 & \hfill a_{\SSS 68} \leq M < a_{\SSS 66} \\ [2ex] 
\DS C + a_{\SSS 65} \left( M - a_{\SSS 67} \right) & \hfill  
a_{\SSS 67} < M \hspace{2.5em}
\end{array} \right. \quad \left( \ref{e:Ralpha} \rmn{a} \right) \]
where $B = {\alpha}_{\SSS R} \left( M = a_{\SSS 66} \right)$ and 
$C = {\alpha}_{\SSS R} \left( M = a_{\SSS 67} \right)$.

The radius $\beta$ coefficient is approximated by 
${\beta}_{\SSS R} = {\beta}_{\SSS R}' - 1$, where 
\beq\label{e:Rbeta}
{\beta}_{\SSS R}' = \frac{a_{\SSS 69} M^{3.5}}{a_{\SSS 70} + M^{{\TS a}_{71}}} 
\qquad 2.0 \leq M \leq 16.0 \, ,
\eeq
with $a_{\SSS 71} \approx 3.45$, and then 
\[ {\beta}_{\SSS R}' = \left\{ \begin{array} {l@{\quad}l} 
\DS 1.06 & \hfill M \leq 1.0 \\ [2ex] 
\multicolumn{2}{l}{
\DS 1.06 + \left( a_{\SSS 72} - 1.06 \right) \left( M - 1.0 \right) /
\left( a_{\SSS 74} - 1.06 \right) 
} \\ [1ex] 
 & \hfill 1.0 < M < a_{\SSS 74} \\ [2ex] 
\multicolumn{2}{l}{
\DS a_{\SSS 72} + \left( B - a_{\SSS 72} \right) \left( M - a_{\SSS 74} \right) 
/ \left( 2.0 - a_{\SSS 74} \right) 
} \\ [1ex] 
 & \hfill a_{\SSS 74} \leq M < 2.0 \\ [2ex] 
\DS C + a_{\SSS 73} \left( M - 16.0 \right) & \hfill 16.0 < M \hspace{2.5em}
\end{array} \right. \quad \left( \ref{e:Rbeta} \rmn{a} \right) \]
where $B = {\beta}_{\SSS R}' \left( M = 2.0 \right)$,  
$C = {\beta}_{\SSS R}' \left( M = 16.0 \right)$.   

If $M > a_{\SSS 75} + 0.1$ then ${\gamma} = 0.0$ where 
$a_{\SSS 75} \approx 1.25$. 
Otherwise 
\beq
\gamma = \left\{ \begin{array} {c@{\quad}l} 
\DS a_{\SSS 76} + a_{\SSS 77} \left( M - a_{\SSS 78} \right)^{{\TS a}_{79}} 
& M \leq 1.0 \\ [1ex] 
\DS B + \left( a_{\SSS 80} - B \right) \left[ \frac{M - 1.0}{a_{\SSS 75} - 1.0} 
\right]^{{\TS a}_{81}} & 1.0 < M \leq a_{\SSS 75} \\ [3ex] 
\DS C - 10.0 \left( M - a_{\SSS 75} \right) C & a_{\SSS 75} < M < 
a_{\SSS 75} + 1.0 
\end{array} \right.
\eeq
where $B = \gamma \left( M = 1.0 \right)$ and $C = a_{\SSS 80}$ unless 
$a_{\SSS 75} \leq 1.0$ when $C = B$.
Note we must always double-check that $\gamma \geq 0.0$. 

Following Tout et al.\ (1997) we note that low-mass MS stars can be 
substantially degenerate below about $0.1 \Msun$ so we take 
\beq
\RMS = \max \left( \RMS , 0.0258 \left( 1.0 + X \right)^{5/3} 
M^{-1/3} \right) 
\eeq
for such stars.

\subsubsection{Hertzsprung gap evolution}

During the HG we define 
\beq
\tau = \frac{t - \tMS}{\tBGB - \tMS } \, . 
\eeq
Then for the luminosity and radius we simply take  
\begin{eqnarray}
L_\rmn{HG} = \LTMS \left( \frac{L_\rmn{EHG}}{\LTMS } \right)^{\tau} \\
R_\rmn{HG} = \RTMS \left( \frac{R_\rmn{EHG}}{\RTMS } \right)^{\tau} \, .
\end{eqnarray}

On the MS we don't consider the core to be dense enough with respect to 
the envelope to actually define a core mass, ie. $M_\rmn{c,MS} = 0.0$.
The core mass at the end of the HG is  
\beq\label{e:mc1hg}
M_\rmn{c,EHG} = \left\{ \begin{array} {l@{\quad}l} 
M_\rmn{c,GB} \left( L = \LBGB \right) & M < \MHeF \\
\McBGB & \MHeF \leq M < \MFGB \\ 
\McHeI & M \geq \MFGB \, , 
\end{array} \right. 
\eeq
where $M_\rmn{c,GB}$, $\McBGB$ and $\McHeI$ will be defined in 
Sections~\ref{s:FGB} and \ref{s:corehb}.
At the beginning of the HG we set $M_\rmn{c,TMS} = \rho M_\rmn{c,EHG}$,  
where 
\beq\label{e:mc2hg}
\rho = \frac{1.586 + M^{5.25}}{2.434 + 1.02 M^{5.25}} \, , 
\eeq
and simply allow the core mass to grow linearly with time so that 
\beq\label{e:mc3hg}
M_\rmn{c,HG} = \left[ \left( 1 - \tau \right) \rho + \tau \right] 
M_\rmn{c,EHG} \, . 
\eeq
If the HG star is losing mass (as described in Section~\ref{s:mloss}) it is 
necessary to take $M_\rmn{c,HG}$ as the maximum of the core mass at the 
previous timestep and the value given by eq.~(\ref{e:mc3hg}). 

\subsection{First giant branch}
\label{s:FGB}

The evolution along the first giant branch (GB) can be modelled, following 
Eggleton, Fitchett \& Tout (1989), using a power-law core mass-luminosity 
relation, 
\beq
L  = \: D M_c^p \, .
\eeq
The evolution is then determined by the growth of the core mass 
as a result of H burning which, in a state of thermal 
equilibrium, is given by
\beq\label{e:Mcdot}
 L \: = \: E X_e \dot{\Mc} \: \Rightarrow \: \dot{\Mc} = A_\rmn{H} L 
\eeq
where
\begin{eqnarray*}
X_e & = & \mbox{ envelope mass fraction of hydrogen, } \\
E & = & \mbox{ the specific energy release and } \\
A_\rmn{H} & = & \mbox{ hydrogen rate constant. }
\end{eqnarray*}
Thus 
\beq
\frac{d M_c}{dt} = A_\rmn{H} D M_c^p
\eeq
which upon integration gives
\beq\label{e:Mcgbt}
M_c = \left[ \left( p - 1 \right) A_\rmn{H} D \left( t_\rmn{inf} - t \right) 
 \right]^{\frac{1}{1 - p}} 
\eeq
or 
\beq
L = D \left[ \left( p - 1 \right) A_\rmn{H} D \left( t_\rmn{inf} - t \right) 
 \right]^{\frac{p}{1 - p}} 
\eeq
so that the time evolution of either $M_c$ or $L$ is given and we can then 
simply find the other from the \Mc-$L$ relation.
Also, when $L = \LBGB$ we have $t = \tBGB$ which defines the integration 
constant 
\beq
t_\rmn{inf} = \tBGB + \frac{1}{A_\rmn{H} D \left( p - 1 \right)} 
\left( \frac{D}{\LBGB} \right)^{\frac{p-1}{p}} \, . 
\eeq

\begin{figure}
\centerline{
\psfig{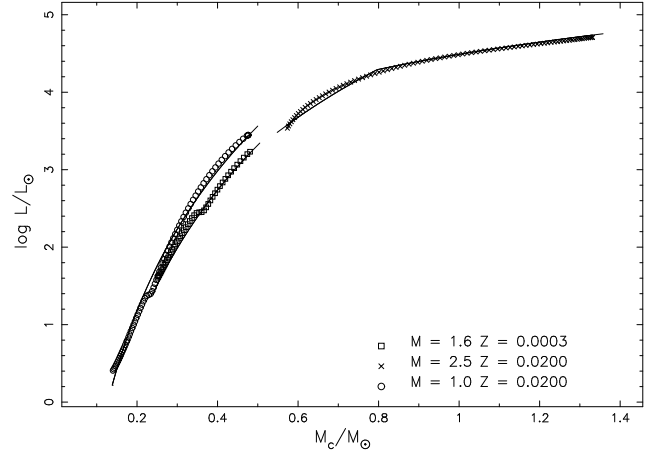}
}
\caption{Relation between core mass and luminosity on the giant branch
showing the fit to points taken from selected detailed models
given by eq.~(\ref{e:McL}) (solid lines).}
\label{f:ffig11}
\end{figure}

Now as noted in Tout et al.\ (1997), the single power-law $L \propto M_c^6$ 
is a good 
approximation to the evolution for small \Mc\ but the relation 
flattens out as \Mc\ approaches the Chandrasekhar mass $\MCh$. 
They expanded the relation to consist of two power-law parts.  
We use an improved form which, albeit somewhat more ad hoc, 
follows much better the actual time evolution along the GB.
Our \Mc-$L$ relation has the form
\beq\label{e:McL}
L = \min \left(B \Mc^q, D \Mc^p \right) \qquad (q < p),
\eeq
so that the first part describes the high-luminosity end and the second the
low-$L$ end of the relation with the two crossing at 
\beq
M_x = \left( \frac{B}{D} \right)^{\frac{1}{p-q}} \, . 
\eeq 
The parameters $B$, $D$, $p$ and $q$ are constants in time for each model 
and indeed are constant in mass for $M < \MHeF$. 
For $M > \MHeF$ it is necessary to introduce a dependence on initial mass 
so that we actually have a \Mc-$L$-$M$ relation. 
The only region in the \Mc-$L$ parameter space where we find that a 
$Z$-dependence is required is in the value of $D$ for $M < \MHeF$.
The parameters are
\[
p = \left\{ \begin{array} {l@{\quad}l} 
6 & M \leq \MHeF \\ 
5 & M \geq 2.5 
\end{array} \right. 
\]
\[
q = \left\{ \begin{array} {l@{\quad}l} 
3 & M \leq \MHeF \\ 
2 & M \geq 2.5 
\end{array} \right. 
\]
\[
B = \max \left( 3 \times 10^4 , 500 + 1.75 \times 10^4 M^{0.6} \right) 
\]
\[
\log D = \left\{ \begin{array} {l@{\quad}l} 
\DS 5.37 + 0.135\zeta \quad \left[ = D_0 \right] & M \leq \MHeF \\ 
\multicolumn{2}{l}{\DS 
\max \left( -1.0 , 0.975 D_0 - 0.18 M , 0.5 D_0 - 0.06 M \right) } \\ 
 & M \geq 2.5 
\end{array} \right. 
\]
with linear interpolation over the transition region, $ \MHeF < M < 2.5$, 
in order to keep the parameters continuous in $M$. 
Thus isochrones constructed with these functions will not give a discontinuity 
on the GB. 
The behaviour of eq.~(\ref{e:McL}) is shown in Fig.~\ref{f:ffig11} as the 
fit to selected model points (note how the relation flattens out as the 
luminosity increases). 

Equation~(\ref{e:Mcgbt}) now becomes 
\beq\label{e:Mcgbt2}
M_\rmn{c,GB} = \left\{ \begin{array} {l@{\quad}l} 
\left[ \left( p - 1 \right) A_\rmn{H} D \left( t_\rmn{inf,1} - t \right) 
\right]^{\frac{1}{1 - p}} & t \leq t_x \\ 
\left[ \left( q - 1 \right) A_\rmn{H} B \left( t_\rmn{inf,2} - t \right) 
\right]^{\frac{1}{1 - q}} & t > t_x 
\end{array} \right. 
\eeq
for $\tBGB \leq t \leq \tHeI$, where 
\begin{eqnarray}
t_\rmn{inf,1} & = & \tBGB + \frac{1}{ \left( p - 1 \right) A_\rmn{H} D} 
\left( \frac{D}{\LBGB } \right)^{\frac{p-1}{p}} \\
t_x & = & t_\rmn{inf,1} - \left( t_\rmn{inf,1} - \tBGB \right) 
\left( \frac{\LBGB }{L_x} \right)^{\frac{p-1}{p}} \\
t_\rmn{inf,2} & = & t_x + \frac{1}{ \left( q - 1 \right) A_\rmn{H} B} 
\left( \frac{B}{L_x} \right)^{\frac{q-1}{q}} \, . \label{e:t2inf} 
\end{eqnarray}
The GB ends at $t = \tHeI$, corresponding to $L = \LHeI$ 
(see Section~\ref{s:corehb}), given by   
\beq
\tHeI = \left\{ \begin{array} {l@{\quad}l} 
t_\rmn{inf,1} - \frac{1}{ \left( p - 1 \right) A_\rmn{H} D} 
\left( \frac{\mbox{$D$}}{\LHeI } \right)^{\frac{p-1}{p}} & \LHeI \leq L_x \\
t_\rmn{inf,2} - \frac{1}{ \left( q - 1 \right) A_\rmn{H} B} 
\left( \frac{\mbox{$B$}}{\LHeI } \right)^{\frac{q-1}{q}} & \LHeI > L_x 
\end{array} \right. \, .
\eeq

The value used for $A_\rmn{H}$ depends on whether we take the PP chain or the 
CNO cycle as the hydrogen burning mechanism with the 
CNO cycle being the most likely on the GB.
Now 
\[ E = {\epsilon}_{CNO} / m \left( He^4 \right) \approx 6.018 \times 10^{18} 
\mbox{ erg/g} \] 
thus
\[ A_\rmn{H} = \left( E X_e \right)^{-1} = 
2.37383 \times 10^{-19} \mbox{ g/erg} \] 
\[ \Rightarrow \quad A_\rmn{H} \approx 1.44 \times 10^{-5} 
\Msun {\Lsun}^{-1} {\rmn{Myr}}^{-1} 
\, , \]
i.e. $\log A_\rmn{H} = -4.84$. 
In practice there are small deviations from thermal equilibrium which 
increase with stellar mass. 
As the value of $A_\rmn{H}$ fixes the rate of evolution on the GB and thus the 
GB timescale it is important for it to be accurate especially if we want  
to use the formulae for population synthesis. 
We find that the detailed models are best represented if we  
introduce a mass dependant $A_\rmn{H}$, ie. $A_{\rmn{H}}'$, where 
\[
\log A_{\rmn{H}}' = \max \left( -4.8 , \min 
\left( -5.7 + 0.8 M , -4.1 + 0.14 M \right) \right) \, .
\]
Some representative values of $A_{\rmn{H}}'$ as a function of 
initial stellar mass 
are shown in Table~1 along with approximate values for 
the GB lifetime and the time taken to reach the GB.
\begin{table}
\begin{center}
\begin{tabular}{cccc}
$M$ & $\log A_{\rmn{H}}'$ & $\frac{\tHeI - \tBGB}{\tBGB}$ & 
$\tBGB$/Myr \\
 & & & \\
1.0 & ${\DS -}4.8$ & $6.4 \times 10^{-2}$ & $10^4$ \\
2.0 & ${\DS -}4.1$ & $2.0 \times 10^{-2}$ & $10^3$ \\
5.0 & ${\DS -}3.4$ & $2.4 \times 10^{-3}$ & $10^2$ \\
\end{tabular}
\end{center}
\caption{A selection of values for the mass dependent hydrogen rate constant 
with approximate timescales also listed.}
\end{table}

Evolution on the GB actually falls into two fairly distinct categories 
depending on whether the initial mass of the star is greater than or 
less than $\MHeF$. 
If $M < \MHeF$ then the star has a degenerate helium core on the GB which grows 
according to the $M_\rmn{c,GB}$ relation derived from eq.~(\ref{e:McL}).
When helium ignites at the tip of the GB it does so degenerately 
resulting in the helium flash. 
However, for IM stars on the GB, $M > \MHeF$, the helium core is 
non-degenerate and the relative time spent on the GB is much shorter and 
thus the models show that $M_\rmn{c,GB}$ is approximately constant from 
the BGB to HeI. 
In this case we still use all the above equations to calculate the timescales
and the luminosity evolution but the corresponding value of $M_\rmn{c}$
is a dummy variable.
The actual core mass at the BGB is given by a mass-dependant formula 
\beq\label{e:Mcbgb}
\McBGB = \min \left( 0.95 \McBAGB , \left( C + c_1 M^{c_2} 
\right)^{\frac{1}{4}} \right) 
\eeq
with $C = \Mc(\LBGB(\MHeF))^4 - c_1 \MHeF^{c_2}$, ensuring the formula is
continuous with the \Mc-$L$ relation at $M = \MHeF$, and $\McBAGB$ given by 
eq.~(\ref{e:McBAGB}).  
The constants $c_1 = 9.20925\,10^{-5}$ and $c_2 = 5.402216$ are  
independent of $Z$, so that for large enough $M$ we have 
$\McBGB \approx 0.098 M^{1.35}$ independent of $Z$. 
Thus on the GB we simply take  
\beq
M_\rmn{c,GB} = \McBGB + \left( \McHeI - \McBGB \right) \tau \quad M > \MHeF 
\eeq
with
\[ \tau = \frac{t - \tBGB}{\tHeI - \tBGB} \]
to account for the small growth of the non-degenerate core while 
$M_\rmn{c,GB}$ is given by eq.~(\ref{e:Mcgbt2}) for $M < \MHeF$. 
$\McHeI$ is described in Section~\ref{s:corehb}. 

\begin{figure}
\centerline{
\psfig{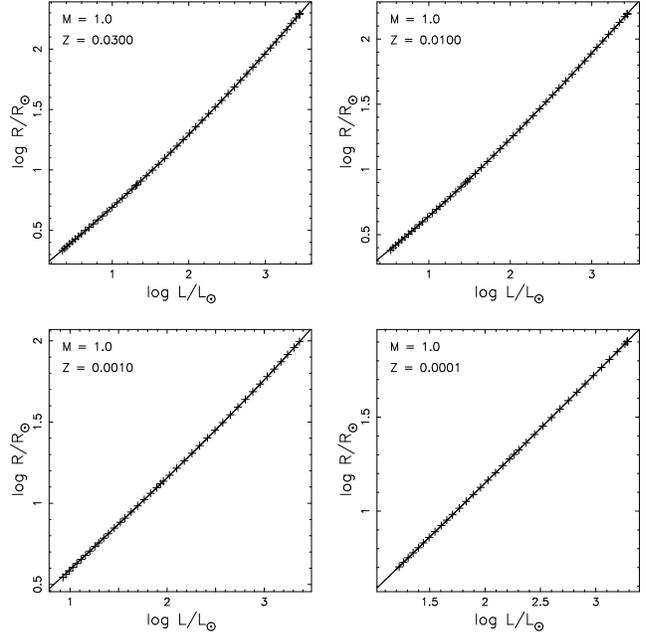}
}
\caption{Relation between radius and luminosity on the giant branch as given by
eq.~(\ref{e:Rgb}) (solid line) and from the detailed models (crosses)
for $1.0 \Msun$ at various metallicities.} 
\label{f:ffig12}
\end{figure}

Furthermore, as giants have a deep convective envelope and thus lie close to 
the Hayashi track, we can find the radius as a function of $L$ and $M$,
\beq\label{e:Rgb}
\Rgb = A \left( L^{{\TS b}_1} + b_2 L^{{\TS b}_3} \right) 
\eeq
where 
\[ A = \min \left( b_4 M^{-{\TS b}_5} , b_6 M^{-{\TS b}_7} \right) \]
and $b_1 \approx 0.4$, $b_2 \approx 0.5$ and $b_3 \approx 0.7$.
A useful quantity is the exponent $x$ to which $R$ depends on $M$ at 
constant $L$, $\Rgb \propto M^{-x}$. 
Thus we also fit $x$ across the entire mass range by 
$A = b M^{-x}$, ie. a hybrid of $b_5$ and $b_7$, to give 
\beq
x = 0.30406 + 0.0805 \zeta + 0.0897 {\zeta}^2 + 0.0878 {\zeta}^3 
+ 0.0222 {\zeta}^4 
\eeq
so that it can be used if required. 
Thus for $Z = 0.02$, as an example, we have 
\beq
\Rgb \approx 1.1 M^{-0.3} \left( L^{0.4} + 0.383 L^{0.76} \right) \, .
\eeq
Figure~\ref{f:ffig12} exhibits the accuracy of eq.~(\ref{e:Rgb}) for solar 
mass models of various metallicity. 

\subsection{Core helium burning}
\label{s:corehb}

The behaviour of stellar models in the HRD during CHeB 
is fairly complicated and depends strongly on the mass and metallicity. 
For LM stars, He ignites at the top of the GB and CHeB corresponds to the
horizontal branch (including the often observed red clump); 
the transition between the He flash and the start of steady CHeB at the 
ZAHB is very rapid and we take it to be instantaneous.
For IM stars, CHeB can
be roughly divided in two phases, descent along the GB to a minimum
luminosity, followed by a blue loop excursion to higher \Teff\ connecting
back up to the base of the AGB (BAGB). However, not all IM stars exhibit a
blue loop, in some cases staying close to the GB throughout CHeB
(the so-called `failed blue loop').
Sometimes the blue loop is also followed by another period of CHeB on the GB
but this is usually much shorter than the first phase and we choose to
ignore it.
For HM stars, He ignites in the HG and CHeB also consists of two phases, 
a blue phase before reaching the GB followed by a red (super)giant phase.

For the purpose of modelling, we define the blue phase of CHeB as that part
which is not spent on the giant branch. This means that the position in the
H-R diagram during the blue phase can in fact be quite red, e.g.\ it includes
the red clump and failed blue loops.  By definition, for the LM regime
the whole of CHeB is blue. For IM stars, the blue phase comes after the RG
phase, while for HM stars it precedes the RG phase.

The transition between the LM and IM star regime occurs over a small mass
range (a few times 0.1\Msun), but it can be modelled in a continuous way with
a factor of the form $1 + \alpha \exp15(M\!-\!\MHeF)$ in the LM formulae (see
below). With $\alpha$ of order unity, this factor can be neglected if $M \ll
\MHeF$.
We also require continuity of LM CHeB stars with naked He stars when the
envelope mass goes to zero.
The formulae are also continuous between IM and HM stars for $Z \le 0.002$.
For higher $Z$, however, there is a discontinuity in the CHeB formulae at $M
= \MFGB$, because the transition becomes too complicated to model continuously
while keeping the formulae simple.

The luminosity at helium ignition is approximated by
\beq\label{e:lhei}
\LHeI = \left\{ \begin{array}{l@{\qquad}l}
\DS \frac{b_9 M^{{\TS b}_{10}}}{1 + {\alpha}_1 \exp15(M\!-\!\MHeF)} &
M < \MHeF \\[3ex]
\DS \frac{b_{11} + b_{12} M^{3.8}}{b_{13} + M^2} & M \ge \MHeF
\end{array} \right.
\eeq
with ${\alpha}_1 = [b_9 \MHeF^{{\TS b}_{10}} - \LHeI(\MHeF)]/\LHeI(\MHeF)$.
The radius at
He ignition is $\RHeI = \Rgb(M,\LHeI)$ for $M \le \MFGB$, and $\RHeI = \RmHe$
for $M \ge \max (\MFGB, 12.0)$,
with \RmHe\ given by eq.~(\ref{e:RmHe}) below.
If $\MFGB < M < 12.0$, we take
\beq
\RHeI = \RmHe \left(\frac{\Rgb(\LHeI)}{\RmHe}\right)^\mu, \quad
\mu = \frac{\log(M/12.0)}{\log(\MFGB/12.0)}. 
\eeq

The minimum luminosity during CHeB for IM stars, reached at the start of the
blue phase, is given by
\beq
\LminHe = \LHeI \frac{b_{14} + c M^{{\TS b}_{15} + 0.1}}{b_{16} + 
M^{{\TS b}_{15}}}
\eeq
with 
\[ c = \frac{b_{17}}{\MFGB^{0.1}} + \frac{b_{16} b_{17} - b_{14}}
{\MFGB^{{\TS b}_{15} + 0.1}}, \]
so that $\LminHe = b_{17} \LHeI$ at $M = \MFGB$. Continuity with HM stars, for
which there is no minimum luminosity, is achieved by taking $b_{17} = 1$ for
$Z\le 0.002$ (but $b_{17} < 1$ for $Z > 0.002$). The radius at this point is
$\Rgb(M,\LminHe)$.

For LM stars the ZAHB luminosity \LZAHB\ takes the place of \LminHe.
To model the ZAHB continuously both with the minimum luminosity point
at $M = \MHeF$ and with the naked He star ZAMS (see Section~\ref{s:naked}) 
for vanishing envelope mass
($M = \Mc$), the ZAHB position must depend on $\Mc$ as well as $M$.
We define
\beq 
\mu = \frac{M - \Mc}{\MHeF - \Mc}, 
\eeq
so that $0 \le \mu \le 1$, and then take
\begin{eqnarray}
\lefteqn{\LZAHB = \LZHe(\Mc) \: +} \nonumber \\
& & \frac{1 + b_{20}}{1 + b_{20} \mu^{1.6479}} \cdot
    \frac{b_{18} \mu^{{\TS b}_{19}}}{1 + {\alpha}_2 \exp15(M\!-\!\MHeF)};
\end{eqnarray}
\[
{\alpha}_2 = \frac{b_{18} + \LZHe(\Mc) - \LminHe(\MHeF)}{\LminHe(\MHeF) -
\LZHe(\Mc)}, 
\]
where $\LZHe$ is defined by eq.~(\ref{e:LZHe}). 
(Note that this ${\alpha}_2$ is not a constant but depends on \Mc.)
For the ZAHB radius we take
\beq
\RZAHB = (1 - f)\RZHe(\Mc) + f \Rgb(\LZAHB);
\eeq
\[
f = \frac{(1.0 + b_{21})\mu^{{\TS b}_{22}}}{1.0 + b_{21}\mu^{{\TS b}_{23}}}.
\]
This formula ensures, apart from continuity at both ends, that \RZAHB\ is
always smaller than the GB radius at \LZAHB.

The minimum radius during the blue loop is approximated by
\beq\label{e:RmHe}
\RmHe = \frac{b_{24} M + (b_{25} M)^{{\TS b}_{26}} M^{{\TS b}_{28}}}
{b_{27} + M^{{\TS b}_{28}}} \qquad M \ge \MHeF. 
\eeq
Then for $M < \MHeF$, we simply take \[ \RmHe = \Rgb(\LZAHB)
\left[\frac{\RmHe(\MHeF)}{\Rgb(\LZAHB(\MHeF))}\right]^\mu \]
to keep \RmHe\ continuous.

The luminosity at the base of the AGB (or the end of CHeB) is given by
\beq
\LBAGB = \left\{ \begin{array}{l@{\qquad}l}
\DS \frac{b_{29} M^{{\TS b}_{30}}}{1 + {\alpha}_3 \exp15(M\!-\!\MHeF)} &
M < \MHeF \\[3ex]
\DS \frac{b_{31} + b_{32} M^{{\TS b}_{33} + 1.8}}{b_{34} + M^{{\TS b}_{33}}} 
& M \ge \MHeF \end{array} \right. 
\eeq
with ${\alpha}_3 = [b_{29} \MHeF^{{\TS b}_{30}} - \LBAGB(\MHeF)]/\LBAGB(\MHeF)$.
The radius at the BAGB is simply $\Ragb(M,\LBAGB)$, 
as given by eq.~(\ref{e:Ragb}). 

The lifetime of CHeB is given by
\beq
\tHe = \left\{ \begin{array}{l@{\qquad}l}
\multicolumn{2}{l}{\DS \left[ b_{39} + \{\tHeMS(\Mc) - b_{39}\}
    \left(1 - \mu\right)^{{\TS b}_{40}} \right]} \\ [1ex]
\times [1 + {\alpha}_4 \exp15(M\!-\!\MHeF)] & M < \MHeF \\[2ex]
\DS \tBGB \frac{b_{41} M^{{\TS b}_{42}} + b_{43} M^5}{b_{44} + M^5} & 
M \ge \MHeF
\end{array} \right.
\eeq
with ${\alpha}_4 = [\tHe(\MHeF) - b_{39}]/b_{39}$.
The term involving \tHeMS(\Mc) ensures continuity with
the lifetime of a naked He star with $M = \Mc$ as the envelope mass vanishes.
The lifetime of the blue phase of CHeB relative to \tHe\ depends in a
complicated way on $M$ and $Z$, it is roughly approximated by
\beq\label{e:taubl}
\tbl = \left\{ \begin{array}{l@{\qquad}l}
1 & M < \MHeF \\[2ex]
\multicolumn{2}{l}{\DS b_{45} \left(\frac{M}{\MFGB}\right)^{0.414} +
{\alpha}_{\SSS bl}
\left(\log\frac{M}{\MFGB}\right)^{{\TS b}_{46}}} \\[2ex]
 & \MHeF \le M \le \MFGB \\[2ex]
\DS (1 - b_{47}) \frac{f_\rmn{bl}(M)}{f_\rmn{bl}(\MFGB)} & M > \MFGB
\end{array} \right.
\eeq
truncated if necessary to give $0 \le \tbl \le 1$,
where \[ {\alpha}_{\SSS bl} =
\Big[1 - b_{45} \left(\frac{\MHeF}{\MFGB}\right)^{0.414}\Big]
\Big[\log\frac{\MHeF}{\MFGB}\Big]^{-{\TS b}_{46}} \]
and \[ f_\rmn{bl}(M) = M^{{\TS b}_{48}} \left[1 -
\frac{\RmHe(M)}{\Ragb(\LHeI(M))}\right]^{{\TS b}_{49}}.\]
The second term in the IM part of eq.~(\ref{e:taubl}) with
${\alpha}_{\SSS bl}$ as
defined ensures that $\tbl = 1$ at $M = \MHeF$.  By taking $b_{45} = 1$ for
$Z\le 0.002$ we also have $\tbl = 1$ at $M = \MFGB$.  The HM part also yields
$\tbl = 1$ at $M = \MFGB$ for $Z\le 0.002$, so that the transition is
continuous for low $Z$. For $Z > 0.002$ the transition is regretably
discontinuous.  Finally, the radius dependence of $f_\rmn{bl}$ ensures that
$\tbl = 0$ at the same mass where $\RmHe = \Ragb(\LHeI)$, i.e.\ where the
blue phase vanishes.

During CHeB, we use the relative age $\tau = (t - \tHeI)/\tHe$ which takes
values between 0 and 1.
We define $\tau_x$ as the relative age at the start of the blue phase of
CHeB, and $L_x$ and $R_x$ are the luminosity and radius at this epoch. Hence,
$\tau_x = 0$ for both the LM and HM regime, and $\tau_x = 1 - \tbl$ for IM
stars,
\beq
L_x = \left\{ \begin{array}{l@{\qquad}l}
\DS \LZAHB & M < \MHeF \\
\DS \LminHe & \MHeF \le M < \MFGB \\
\DS \LHeI & M \ge \MFGB
\end{array} \right.
\eeq
and
\beq
R_x = \left\{ \begin{array}{l@{\qquad}l}
\DS \RZAHB & M < \MHeF \\
\DS \Rgb(\LminHe) & \MHeF \le M < \MFGB \\
\DS \RHeI & M \ge \MFGB
\end{array} \right.
\eeq
Then the luminosity during CHeB is modelled as
\beq\label{e:LCHeB}
L = \left\{ \begin{array}{l@{\qquad}l}
\DS L_x \left(\frac{\LBAGB}{L_x}\right)^\lambda & \tau_x \le \tau \le 1 \\[2ex]
\DS L_x \left(\frac{\LHeI}{L_x}\right)^{\lambda'} & 0 \le \tau < \tau_x
\end{array} \right.
\eeq
where
\beq
\lambda = \left(\frac{\tau - \tau_x}{1 - \tau_x}\right)^\xi; \quad
\xi = \min(2.5, \max (0.4, \RmHe/R_x)),
\eeq
\beq
\lambda' = \left(\frac{\tau_x - \tau}{\tau_x}\right)^3.
\eeq

The \emphh{actual} minimum radius during CHeB is
$\Rmin = \min (\RmHe, R_x)$,
because eq.~(\ref{e:RmHe}) for \RmHe\ can give a value that
is greater than $R_x$ (this property is used, however, to compute $\xi$
above).  Furthermore, we define $\tau_y$ as the relative age at the end of
the blue phase of CHeB, and $L_y$ and $R_y$ as the luminosity and radius at
$\tau = \tau_y$. Hence, $\tau_y = 1$ for LM and IM stars and $\tau_y = \tbl$
for IM stars. $L_y$ is given by eq.~(\ref{e:LCHeB}) ($L_y = \LBAGB$ for $M
\le \MFGB$), and $R_y = \Ragb(L_y)$.  The radius during CHeB is modelled
as
\beq\label{e:RCHeB}
R = \left\{ \begin{array}{l@{\qquad}l}
\DS \Rgb(M,L) & 0 \le \tau < \tau_x \\
\DS \Ragb(M,L) & \tau_y < \tau \le 1 \\
\DS \Rmin \exp(|\rho|^3) & \tau_x \le \tau \le \tau_y
\end{array} \right.
\eeq
where
\beq
\rho = \left(\ln\frac{R_y}{\Rmin}\right)^{\!\frac{1}{3}} \!
  \Big(\frac{\tau - \tau_x}{\tau_y - \tau_x}\Big)
  - \left(\ln\frac{R_x}{\Rmin}\right)^{\!\frac{1}{3}} \!
  \Big(\frac{\tau_y - \tau}{\tau_y - \tau_x}\Big).
\eeq

The core mass \McHeI\ at He ignition is given by the \Mc-$L$ relation for LM
stars, while for $M \ge \MHeF$ the same formula can be used as for the BGB
core mass (eq.~\ref{e:Mcbgb}) replacing \Mc(\LBGB(\MHeF)) with
\Mc(\LHeI(\MHeF)) to ensure continuous transition at $M = \MHeF$. For $M >
3\Msun$, \McHeI\ is nearly equal to \McBGB.
The core mass at the BAGB 
point is approximated by
\beq\label{e:McBAGB}
\McBAGB = (b_{36} M^{{\TS b}_{37}} + b_{38})^\frac{1}{4}
\eeq
where $b_{36} \approx 4.36 \times 10^{-4}$, $b_{37} \approx 5.22$ and
$b_{38} \approx 6.84 \times 10^{-2}$.
In between the core mass is taken to simply increase linearly with time 
\beq
\Mc = (1 - \tau) \McHeI + \tau \McBAGB.
\eeq

\subsection{Asymptotic giant branch}
\label{s:agbev}

During the EAGB, when the H-burning shell 
is extinct, the (H-exhausted) core
mass \Mche\ (which we have been calling \Mc\ so far because it was the only 
significant core) stays constant at the value \McBAGB. 
Within the H-exhausted core a degenerate carbon-oxygen core, \Mcco, has formed 
and begins to grow. 
At a time corresponding to the second dredge-up phase the growing \Mcco\  
catches the H-exhausted core and the TPAGB begins.  
From then on \Mcco\ and \Mche\ are equal and grow at the same rate 
(we neglect the mass, about $0.01 \Msun$, of the thin helium layer between the 
two burning shells).  

So on the EAGB we set  
\[ \Mc = \Mche = \McBAGB \, . \]
Inside this core, \Mcco\ grows 
by He-shell burning, at a rate dictated by the \Mc-$L$ relation.
Thus we can compute the evolution of \Mcco\ and $L$ in the same way as was  
done for GB stars using 
eqs.~(\ref{e:McL}) and (\ref{e:Mcgbt2}) with \Mc\ replaced by \Mcco, 
 $\tBGB$ replaced by $\tBAGB$ ($ = \tHeI + \tHe$) and $\LBGB$ 
replaced by $\LBAGB$.  
We also need to replace $A_\rmn{H}$  
with the value appropriate for He burning, $A_{\rmn{He}}$. 
The detailed models (Pols et al.\ 1998) on the EAGB show that the carbon-oxygen 
core is composed of 20\% carbon and 80\% oxygen by mass so for every 4 carbon 
atoms produced by the triple-$\alpha$ reaction, 3 will capture an $\alpha$ 
particle and be converted to oxygen.  
Thus 
\[ E = \frac{{\epsilon}_{3 \alpha} + 0.75 {\epsilon}_{C \alpha}}{15 \, m \left( 
H \right)} \approx 8.09 \times 10^{17} \mbox{ erg/g} \]  
so that 
\beq\label{e:herate}
A_\rmn{He} = \left( E X_\rmn{He} \right)^{-1} = 7.66 \times 10^{-5} 
\Msun {\Lsun}^{-1} {\rmn{Myr}}^{-1} 
\eeq
using $X_\rmn{He} \approx 0.98$.
Although massive stars ($M \ga 8$) do not actually follow a \Mc-$L$ relation
for the CO core, by making the proper (ad hoc) assumptions about the
constants in the relation, we can still effectively model their evolution in
the same way as for true AGB stars.

As already mentioned, the EAGB ends when the the growing CO-core reaches the 
H-exhausted core. 
If $0.8 < \McBAGB < 2.25$, the star
will undergo a second dredge-up phase at the end of the EAGB phase. 
During this second dredge-up the core mass is reduced to
\beq
\McDU = 0.44\McBAGB + 0.448.
\eeq
We assume that the second dredge-up takes place instantaneously at the
moment when \Mcco\ reaches the value \McDU, so that also $\Mcco = \Mc$ at that
point (but note that there is then a sudden discontinuity in $\Mc = \Mche$).
Similarly, for $\McBAGB \le 0.8$, the EAGB ends when \Mcco\ reaches 
\Mche\ without a second dredge-up, ie. $\McDU = \McBAGB$.
Stars with $\McBAGB > 2.25$ do not undergo second dredge-up, as they 
can ignite carbon non-degenerately, and their
evolution terminates before they ever reach the TPAGB.  

To determine when the transition from EAGB to TPAGB occurs we can simply  
insert \McDU\ into the \Mc-$L$ relation to find 
$L_\rmn{{\SSS DU}}$. 
Then we calculate  
\beq
t_\rmn{{\SSS DU}} = \left\{ \begin{array} {l@{\quad}l} 
t_\rmn{inf,1} - \frac{1}{ \left( p - 1 \right) A_\rmn{He} D} 
\left( \frac{D}{L_\rmn{{\SSS DU}}} \right)^{\frac{p-1}{p}} & L_\rmn{{\SSS DU}} 
\leq L_x \\
t_\rmn{inf,2} - \frac{1}{ \left( q - 1 \right) A_\rmn{He} B} 
\left( \frac{B}{L_\rmn{{\SSS DU}}} \right)^{\frac{q-1}{q}} & L_\rmn{{\SSS DU}} 
> L_x 
\end{array} \right. \, .
\eeq
Thus if $t > t_\rmn{{\SSS DU}}$ the TPAGB has begun and the H-exhausted  
and He-exhausted cores grow together as a common core. 
Once again the \Mc-$L$ relation is obeyed and once again we can use it in 
the same way as we did for GB stars if we replace $\tBGB$ by 
$t_\rmn{{\SSS DU}}$ and $\LBGB$ by $L_\rmn{{\SSS DU}}$. 
As we have both hydrogen and helium shell burning in operation then we 
must also replace $A_\rmn{H}$ by an effective combined rate $A_\rmn{H,He}$ 
where 
\beq
A_\rmn{H,He} = \frac{A_\rmn{H} A_\rmn{He}}
{A_\rmn{H} + A_\rmn{He}} \simeq 1.27 \times 10^{-5} 
\Msun {\Lsun}^{-1} {\rmn{Myr}}^{-1} \, .
\eeq 
There is however an added complication that it is possible for  
$L_\rmn{{\SSS DU}} > L_x$. 
In this case $t_\rmn{inf,1}$ and $t_x$ are not needed and $t_\rmn{inf,2}$ 
is given by 
\beq
t_\rmn{inf,2} = t_\rmn{{\SSS DU}} + \frac{1}{ \left( q - 1 \right) A_\rmn{H,He}
 B}
\left( \frac{B}{L_\rmn{{\SSS DU}}} \right)^{\frac{q-1}{q}} \, .
\eeq
In this way the $L$ evolution (and thus the $R$ evolution) remains  
continuous through the second dredge-up.

\begin{figure}
\centerline{
\psfig{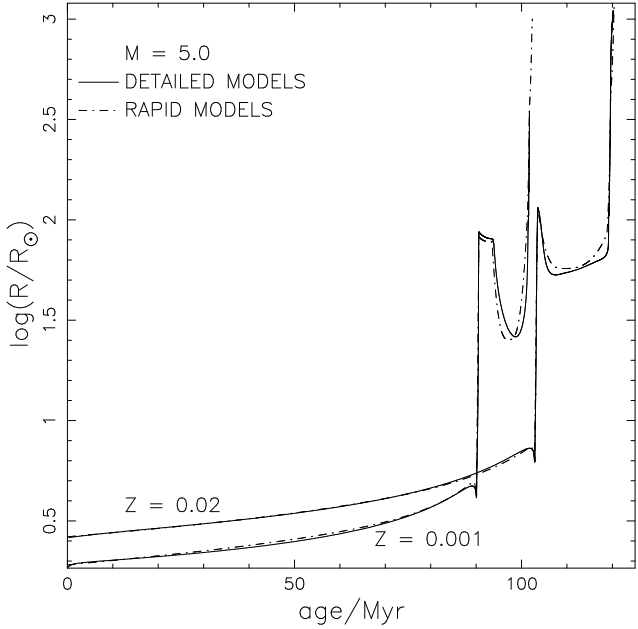}
}
\caption{Radius evolution from the ZAMS to the end of the AGB for a
$5.0 \Msun$ star, for metallicities 0.001 and 0.02, showing the detailed
model points (solid lines) and the fitted tracks (dash-dot lines).}
\label{f:ffig13}
\end{figure}

On the TPAGB we do not model the thermal pulses individually, but we do take
into account the most important effect of the thermally pulsing behaviour
on the long-term evolution, namely that of third dredge-ups.
During each interpulse
period, the He core grows steadily, but during the thermal pulse itself the
convective envelope reaches inwards and takes back part of the mass
previously eaten up by the core.
The fraction of this mass is denoted by $\lambda$.
Frost (1997) shows that models with $4 \leq M  \leq 6$ and $0.004 \leq Z
\leq 0.02$ have similar overall behaviour in $\lambda$ where $\lambda$
increases quickly and reaches approximately 0.9 after about 5 pulses at which 
it stays nearly constant for the remaining pulses.
For lower-mass stars there is no evidence for such a high $\lambda$ with a
value of 0.3 more likely for models of approximately solar mass and then a
steady increase of $\lambda$ with $M$ to reach ${\lambda}_\rmn{max} \approx 
0.9$ before $M = 4$ (Lattanzio 1989; 
Karakas et al.\ 1999)  
Thus we simply take $\lambda$ as constant for each $M$ without any $Z$
dependence,
\beq
\lambda = \min (0.9, 0.3 + 0.001 M^5).
\eeq
Hence, the secular growth of the core mass is reduced with respect to that
given by the \Mc-$L$ relation by a fraction $\lambda$. 
On the other hand, detailed calculations show that the luminosity 
evolution with time follows the same relation as without third dredge-up 
(Frost 1997), ie. it keeps following eqs.~(\ref{e:McL}) and (\ref{e:Mcgbt2}) 
as if \Mc\ were not reduced by dredge-up. 
In other words, the \Mc-$L$ relation is no longer satisfied in the presence
of third dredge-up, but we can use it nevertheless to compute the evolution
of $L$, while \Mc\ is modified as follows:
\[
\Mc = \McDU + (1 - \lambda)(\Mc' - \McDU),
\]
where $\Mc'$ is from the \Mc-$L$ relationship, with no dredge-up, 
and \McDU\ is the value of \Mc\ at the start of the TPAGB.

The radius evolution is very similar to that of the GB, as the stars still 
have a deep convective envelope, but with some slight modifications. 
The basic formula is the same, 
\beq\label{e:Ragb}
\Ragb = A \left( L^{{\TS b}_1} + b_2 L^{{\TS b}_{50}} \right) 
\eeq
where indeed $b_1$ and $b_2$ are exactly the same as for $\Rgb$ and 
$b_{50} = b_{55} b_3$ for $M \geq \MHeF$. 
Also for $M \geq \MHeF$
\[ A = \min \left( b_{51} M^{-{\TS b}_{52}} , b_{53} M^{-{\TS b}_{54}} \right) 
\]
which gives
\[ \Ragb = 1.125 M^{-0.33} \left( L^{0.4} + 0.383 L^{0.76} \right), \] 
as an example, for $Z = 0.02$.
For $M < \MHeF$ the behaviour is slightly altered so we take   
\begin{eqnarray*}
b_{50} & = & b_3 \\ 
A & = & b_{56} + b_{57} M 
\end{eqnarray*}
for $M \leq \MHeF - 0.2$ and linear interpolation between the bounding 
values for $\MHeF - 0.2 < M < \MHeF$, which means that for $M = 1.0$ and 
$Z = 0.02$ the relation gives  
\[ \Ragb \approx 0.95 \left( L^{0.4} + 0.383 L^{0.74} \right) \, . \] 

In Figure~\ref{f:ffig13} we show the radius evolution of a $5.0 \Msun$ star, 
for $Z= 0.001$ and $Z = 0.02$, from the ZAMS to the end of the AGB, from both 
the rapid evolution formulae and the detailed models. 
The AGB phase of the evolution is recognised by the sharp increase in radius 
following the phase of decreasing radius during the CHeB blue loop. 
An accurate fit to the AGB radius is required if the formulae are to be used 
in conjunction with binary evolution where factors such as Roche-lobe 
overflow and tidal circularisation come into play. 
In actual fact Fig.~\ref{f:ffig13} shows that we acheive an accurate fit 
for all phases of the evolution.

\begin{figure}
\centerline{
\psfig{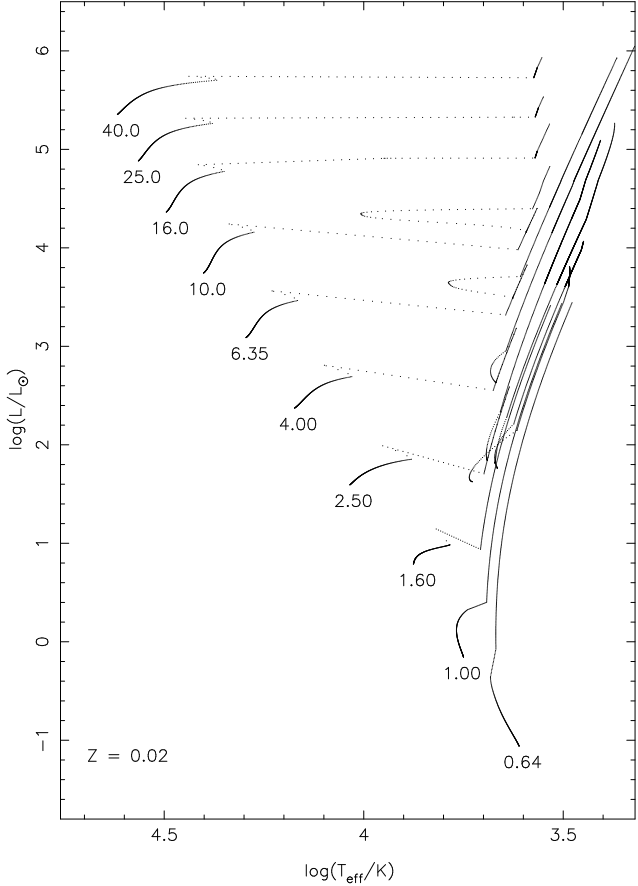}
}
\caption{Same as Fig.~\ref{f:popi} but tracks are from the evolution
formulae.}
\label{f:spopi}
\end{figure}

\begin{figure}
\centerline{
\psfig{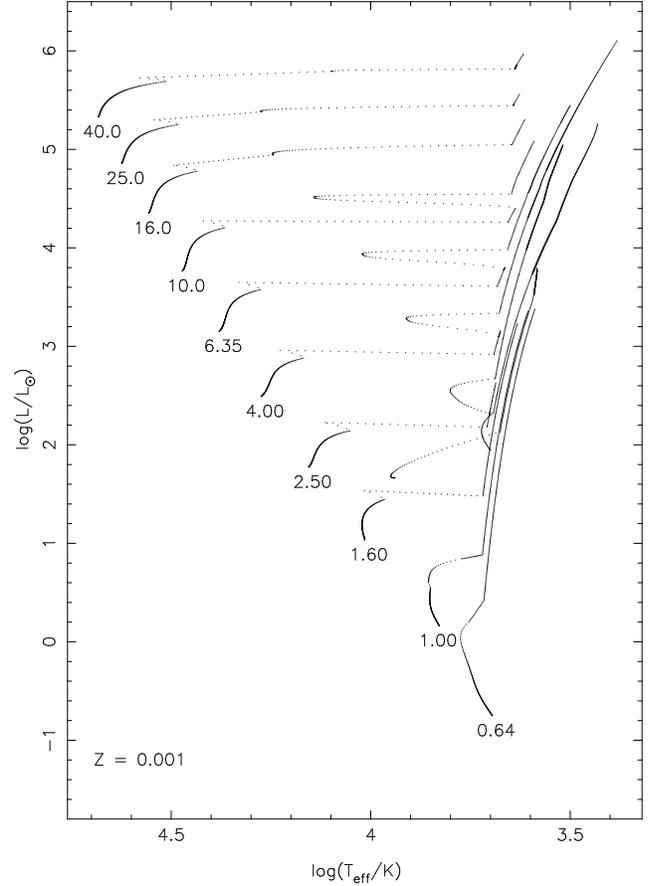}
}
\caption{Same as Fig.~\ref{f:popii} but tracks are from the evolution
formulae.}
\label{f:spopii}
\end{figure}

We have now described formulae which cover all phases of the evolution 
covered by the detailed grid of stellar models. 
Figures~\ref{f:spopi} and \ref{f:spopii} show synthetic HRDs derived from the 
formulae and are designed to be direct comparisons to Figs.~\ref{f:popi} and 
\ref{f:popii} respectively. 
The excellent performance of the fitting formulae is clearly evident. 

\section{Final stages and remnants}
\label{s:finstg}

The AGB evolution is terminated, if not by complete loss of the envelope,
when the CO-core mass reaches a maximum value given by
\beq\label{e:Mcsn}
\McSN = \max (\MCh, 0.773\McBAGB - 0.35).
\eeq
When this maximum core mass is reached before the envelope is lost, a
\emphh{supernova} explosion is assumed to take place. 
For stars with $\McBAGB \le 2.25$, this should occur during the TPAGB phase. 
In practice mass loss will prevent it from doing so in most 
cases of single star evolution, but it may occur as a consequence of 
binary evolution.
For such stars, we make a further distinction based on whether 
\McBAGB\ exceeds 1.6\Msun.
For $\McBAGB < 1.6$, when the CO-core mass reaches \MCh\ carbon
ignites in a degenerate flash, leading to a thermonuclear explosion.
It is uncertain whether we should expect this to occur for normal SSE but 
if it does then the supernova would be something like ``type IIa'' 
(Ia + hydrogen) 
and we assume that such a supernova leaves no stellar remnant.

For $1.6 \le \McBAGB \le 2.25$, the detailed models show that carbon ignites
off-centre under semi-degenerate conditions when $\Mcco \ga 1.08$ (Pols et
al.\ 1998). 
Carbon burning is expected to lead to the formation of a
degenerate ONe-core (Nomoto 1984), while the star continues its evolution up
the AGB. 
When the core mass reaches \MCh, the ONe-core collapses owing to
electron capture on Mg$^{24}$ nuclei. 
The resulting supernova explosion leaves a neutron star remnant 
(Section~\ref{s:nsremn}).
The limiting \McBAGB\ values of
1.6\Msun\ and 2.25\Msun\ correspond to initial stellar masses denoted
traditionally by the symbols \Mup\ and \Mec, respectively. 
The values of \Mup\ and \Mec\ depend on metallicity 
(see Table~1 of Pols et al.\ 1998), this dependence follows from inverting 
eq.~(\ref{e:McBAGB}) for the values $\McBAGB = 1.6$ and 2.25, respectively.

If the envelope is lost before \Mc\ reaches \McSN\ $(= \MCh)$ on the TPAGB,
the remnant core becomes a white dwarf. 
This will be the case for almost all cases of normal SSE.
For $\McBAGB < 1.6$, this will be a CO white dwarf, for
$\McBAGB \ge 1.6$ it will be a ONe white dwarf (Section~\ref{s:wdremn}).

Stars with $\McBAGB > 2.25$ develop
non-degenerate CO-cores which grow only slightly before undergoing central
carbon burning, rapidly followed by burning of heavier elements.  Here,
\McSN\ is the CO-core mass at which this burning takes place, because the
core mass does not grow significantly after C burning.  Very quickly, an 
Fe-core is formed which collapses owing to photo-disintegration, resulting in a
supernova explosion.  The supernova leaves either a neutron star or, for very
massive stars, a black hole (Section~\ref{s:nsremn}).  
We assume that a black hole 
forms if $\McSN > 7.0$, corresponding to $\McBAGB > 9.52$.

This means that the lowest mass star to produce a NS has an initial mass 
$M_*$ in 
the range $\Mup \leq M_* \leq \Mec$ with the actual value of $M_*$ 
depending greatly on the mass loss rate. 
Observations would tend to suggest that $M_* \approx \Mec$ (Elson et al.\ 1998) 
and indeed we find that with our adopted mass loss rate (Section~\ref{s:mloss}) 
almost all cases of SSE result in WD formation for $M \leq \Mec$.

While most stars have their nuclear burning evolution terminated on the TPAGB 
we must make allowances for cases of enhanced mass loss, e.g. owing to binary 
evolution processes, that result in termination at an earlier nuclear 
burning stage. 
If the star loses its envelope during the HG or GB phases then the star will 
become either a HeWD (Section~\ref{s:wdremn}), if it has a degenerate core 
($M \leq \MHeF$), or a zero-age naked He star (Section~\ref{s:naked}). 
If during CHeB $M = \Mc$ then an evolved naked He star is formed with the 
degree of evolution determined by the amount of central helium already 
burnt. 
Thus the age of the new star is taken to be 
\beq
t = \left( \frac{t' - {\tHeI}'}{{\tHe}'} \right) \tHeMS 
\eeq
where the primes denote times for the original star and $\tHeMS$ is given by 
eq.~(\ref{e:thems}). 
When the envelope is lost during the EAGB so that $\Mche = M$, a naked helium  
giant (Section~\ref{s:naked}) is formed as unburnt helium still remains within 
\Mche\ through which the growing \Mcco\ is eating. 
The age of the new star will be fixed by using $\Mc = \Mcco$ and 
$M = \Mche$ in the HeGB \Mc-$t$ relation (see Section~\ref{s:naked}). 
We note that although naked helium stars are nuclear burning stars, 
ie. not a final state, we still label them as a remnant stage because they 
are the result of mass loss. 
Also, when a WD, NS or BH is formed the age of the star is reset so that the 
remnant begins its evolution at zero-age to allow for cooling 
(Section~\ref{s:remn}).

\subsection{Naked helium stars}
\label{s:naked}

The formulae described in this section are based on detailed stellar 
evolution models for naked helium stars, computed by one of the authors (ORP)
with the same code as used for the stellar models described in 
Section~\ref{s:models}. 
First, a helium ZAMS of homogeneous models in thermal
equilibrium was constructed, with composition $X=0$, $Y=0.98$ and $Z=0.02$.
Starting from this ZAMS, evolution tracks were computed for masses between
0.32 and 10~\Msun\ spaced by approximately 0.1 in $\log M$. 
For masses below 2~\Msun, the tracks were computed until the end of shell He
burning and for $M>2\Msun$, up to or through central carbon burning.
These models will be discussed in more detail in a forthcoming paper 
(Pols 1999, in preparation). 

The following analytic formulae provide an accurate fit to the ZAMS 
luminosity and radius of naked He stars with $Z = 0.02$:
\beq\label{e:LZHe}
\LZHe = \frac{15\,262\, M^{10.25}}{M^9 + 29.54 M^{7.5} + 31.18 M^6 + 
0.0469},
\eeq
\beq\label{e:RZHe}
\RZHe = \frac{0.2391 M^{4.6}}{M^4 + 0.162 M^3 + 0.0065}.
\eeq
The central He-burning lifetime (He MS) is approximated by
\beq\label{e:thems}
\tHeMS = \frac{0.4129 + 18.81 M^4 + 1.853 M^6}{M^{6.5}}.
\eeq
The behaviour of $L$ and $R$ during central He burning can be approximated by
\beq\label{e:LHeMS}
L_\rmn{HeMS} = \LZHe (1 + 0.45\tau + \alpha \tau^2)
\eeq
and
\beq\label{e:RHeMS}
R_\rmn{HeMS} = \RZHe (1 + \beta \tau - \beta \tau^6).
\eeq
where $\tau = t/\tHeMS$ and $t$ is counted from the He ZAMS. 
$\alpha$ and $\beta$ are dependent on
mass, as follows:
\beq
\alpha = \max (0, 0.85 - 0.08 M)
\eeq
and
\beq
\beta = \max (0, 0.4 - 0.22 \log M).
\eeq

The evolution after the He MS is dominated by the growth
of the degenerate C-O core for low-mass stars, and by evolution up to carbon
burning for $M \ga 2$. 
Low-mass He stars follow an approximate core mass-luminosity relation 
(e.g.\ Jeffery 1988), and we compute their evolution making use of this 
relation just as we do for GB stars (Section~\ref{s:FGB}).
For massive He stars, although they do not properly follow such a relation,
an ad hoc \Mc-$L$ relation can be used to also describe their evolution.
The following formula works for the whole mass range:
\beq\label{e:lmche}
L_\rmn{HeGB} = \rmn{MIN} (B \Mc^3, D \Mc^5)
\eeq
with $B = 4.1\times10^4$ and $D = 5.5\times10^4 / (1 + 0.4 M^4)$.
The first term models the `real' \Mc-$L$ relation followed by low-mass He 
stars, while the second, mass-dependent term mimics
the behaviour for high-mass stars.
The evolution of $L$ and \Mc\ with time is obtained from eq.~(\ref{e:lmche})
and the equivalents of eqs.~(\ref{e:Mcgbt2}-\ref{e:t2inf}) with $A_\rmn{H}$ 
replaced by $A_\rmn{He}$ as given by eq.~(\ref{e:herate}), 
$\tBGB$ replaced by $\tHeMS$, and $\LBGB$ replaced by $L_{\rm THe}$. 
$L_\rmn{THe}$ is the value of $L$ at the end of the 
He MS, i.e.\ $L_\rmn{HeMS}$ given by eq.~(\ref{e:LHeMS}) at $\tau=1$.
The post-HeMS radius can be approximated by 
\beq
R_\rmn{HeGB} = \rmn{MIN} (R_1, R_2),
\eeq
\beq
R_1 = \RZHe\!\left( \frac{L}{L_{\rm THe}} \right)^{\!0.2} \! + 
0.02 \left[\exp\!\left( \frac{L}{\lambda} \right) - 
\exp\!\left( \frac{L_{\rm THe}}{\lambda} \right) \right], \eeq\beq
\lambda = 500 \frac{2 + M^5}{M^{2.5}},
\eeq
\beq
R_2 = 0.08 L^{0.75}.
\eeq
The first term of $R_1$ models the modest increase in radius at low
mass and/or $L$, and the second term the very rapid expansion and redward
movement in the HRD for $M \ga 0.8$ once $L$ is large enough.
The star is on what we call the naked helium HG (HeHG) if the radius is 
given by $R_1$. 
The radius $R_2$ mimics the Hayashi track for He stars on the giant branch 
(HeGB).
We make the distinction between HeHG and HeGB stars only because the latter 
have deep convective envelopes and will therefore respond differently
to mass loss.

The final stages of evolution are equivalent to those of normal stars, i.e.\
as discussed in Section~\ref{s:finstg}, 
but with \McBAGB\ replaced by the He-star initial mass $M$ in 
eq.~(\ref{e:Mcsn}) as well as in the discussion that follows it.
If $M<0.7\Msun$, the detailed models show that shell He burning stops before
the whole envelope is converted into C and O.  We mimic this by letting a 
He star become a CO WD when its core mass reaches the value
\beq
\Mcmax = \min (1.45 M - 0.31 , M),
\eeq
as long as $\Mcmax < \McSN$.

\subsection{Stellar remnants}
\label{s:remn}

\subsubsection{White dwarfs}
\label{s:wdremn}

We distinguish between three types of white dwarf, those composed of
He (formed by complete envelope loss of a GB star with $M < \MHeF$, only 
expected in binaries), 
those composed of C and O (formed by envelope loss of a TPAGB star
with $M < \Mup$, see above), and those composed mainly of O and Ne (envelope
loss of a TPAGB star with $\Mup \le M \le \Mec$).
The only distinction we make between CO and ONe white dwarfs is in the way
they react to mass accretion.
If $\MWD + M_\rmn{acc} > \MCh$, after accreting an amount of mass 
$M_\rmn{acc}$, then a CO WD explodes without leaving a 
remnant while an ONe WD leaves a neutron star remnant with mass 
$\MNS = 1.17 + 0.09 \left( \MWD + M_\rmn{acc} \right) $ 
(see later in Section~\ref{s:nsremn}).
The Chandrasekhar mass is given by 
\[ \MCh \approx \left( \frac{5.8}{{\mu}^2_\rmn{e}} \right) \Msun \] 
so it is composition dependent but the mean molecular weight per electron 
is ${\mu}_\rmn{e} \approx 2$, except for low-mass MS stars in cataclysmic 
variables, so we use $\MCh = 1.44$ at all times. 

The luminosity evolution of white dwarfs is modelled using standard cooling
theory (Mestel 1952), see Shapiro \& Teukolsky (1983, pg. 85):
\beq\label{e:Lwd}
L_\rmn{WD} = \frac{635 M Z^{0.4}}{ \left[ A (t + 0.1) \right]^{1.4}},
\eeq
where $t$ is the age since formation and $A$ is the effective baryon 
number for the WD composition. 
For He WDs we have $A = 4$ for CO WDs $A = 15$ and for ONe WDs $A = 17$. 
Eqn.~(\ref{e:Lwd}) is adequate for relatively old WDs. 
The addition of a constant in the factor $(t + 0.1)$ mimics the fact that 
the initial cooling is rather faster than given by Mestel theory, 
as well as ensuring that it doesn't start at infinite $L$, 
so that we effectively start the evolution at a cooling age of $10^5$\,yr. 
Note that the initial cooling of the WD is modelled by the small-envelope 
pertubation functions on the TPAGB (see Section~\ref{s:smenv}). 

The radius of a white dwarf is given by
\beq
\RWD = \max \left( \RNS, 0.0115 \sqrt{
\left( \frac{\MCh}{\MWD} \right)^{2/3} - 
\left( \frac{\MWD}{\MCh} \right)^{2/3} } \right)
\eeq
as in Tout et al.\ (1997).

\subsubsection{Neutron stars and black holes}
\label{s:nsremn}

When a neutron star or black hole is formed in one of the
situations given above, we assume that its gravitational mass is given by
\beq\label{e:Mns}
\MNS = 1.17 + 0.09\McSN,
\eeq
where \McSN\ is the mass of the CO-core at the time of supernova explosion.
With eq.~(\ref{e:Mcsn}), this leads to a minimum NS mass of 1.3\Msun, and
the criterion for BH formation $\McSN > 7.0$ gives a maximum NS mass and
minimum BH mass of 1.8\Msun.  

The NS cooling curve is approximated by assuming that photon emission is the 
dominant energy loss mechanism, which should be true for $t \geq 10^6$ yrs 
(see Shapiro \& Teukolsky, pg. 330):
\beq\label{e:Lns}
L_\rmn{NS} = \frac{0.02 M^{2/3}}{ \left( \max \left( t , 0.1 \right) 
\right)^{2}} \, .
\eeq
The upper limit is calibrated to give $\Teff \approx 2 \times 10^6 K$ which is 
appropriate for the Crab Pulsar and is set constant for the first $10^{5}$ 
yrs to reflect the scatter in the observations of \Teff\ for pulsars with an 
age less than $10^{5}$ yrs. 
Eqn.~(\ref{e:Lns}) also ensures that $L_\rmn{NS} < L_\rmn{WD} $ at all times 
and that neutron stars will cool faster than white dwarfs. 

The radius of a NS is simply set to 10 km, i.e.\ $\RNS = 1.4~10^{-5}$.

We take the black hole radius as the Schwarzschild radius 
\beq
\RBH = \frac{2 G M_\rmn{BH}}{c^2} = 4.24~10^{-6} M_\rmn{BH} \, .
\eeq
The corresponding luminosity of a BH is approximately given by 
\beq
L_\rmn{BH} = \frac{1.6 \times 10^{-50}}{M_\rmn{BH}^2} 
\eeq
(Carr \& Hawking 1974) which will be negligible except for extremely low mass 
objects and thus we actually set  
\beq
L_\rmn{BH} = 10^{-10}
\eeq
to avoid floating point division by zero. 

Note that for all remnants we set $\Mc = M$ for convenience. 

\subsection{Small envelope behaviour and hot subdwarfs}
\label{s:smenv}

In general the equations in Section~\ref{s:fitfor} accurately describe the 
nuclear burning evolution stages as outlined by our grid of detailed models.
However, we also find it necessary to add some pertubation functions 
which alter the radius and luminosity when the envelope becomes small in 
mass, in order to achieve a smooth transition in the HRD towards the position 
of the remnant.
Take, for example, the AGB radius where 
\[ \Ragb \propto M^{-x} \]
so that as $M$ decreases due to mass loss from a stellar wind \Ragb\ will 
increase and the star moves further to the red in the HRD. 
In actual fact, as the envelope mass (\Menv) gets very small, the star 
becomes bluer and moves across the HRD to WD temperatures. 
In the same way we would also expect the luminosity growth rate to  
decrease until the luminosity levels off at some approximately constant 
value for small \Menv.

Thus for any nuclear burning evolution stage where there is a well defined 
core and envelope (i.e. not the MS), we define 
\beq\label{e:muenv}
\mu = \left( \frac{M - \Mc}{M} \right) \min \left( 5.0 , 
\max \left( 1.2 , \left( \frac{L}{L_0} \right)^{\kappa} 
\right) \right) 
\eeq
where $L_0 = 7.0 \times 10^4$, $\kappa = -0.5$ for normal giants and 
\beq
\mu = 5 \left( \frac{\Mcmax - \Mc}{\Mcmax} \right) 
\eeq
for helium giants. 
Then if $\mu < 1.0$ we perturb the luminosity and radius using 
\begin{eqnarray}
L' & = & L_\rmn{c} \left( \frac{L}{L_\rmn{c}} \right)^{s} \\ 
R' & = & R_\rmn{c} \left( \frac{R}{R_\rmn{c}} \right)^{r} 
\end{eqnarray}
where
\begin{eqnarray}
s & = & \frac{\left( 1 + b^3 \right) \left( \mu / b \right)^3}
{1 + \left( \mu / b \right)^3} \\ 
r & = & \frac{\left( 1 + c^3 \right) \left( \mu / c \right)^3 
{\mu}^{0.1/q}}
{1 + \left( \mu / c \right)^3} 
\end{eqnarray}
with
\begin{eqnarray}
b & = & 0.002 \max \left( 1 , \frac{2.5}{M} \right) \\
c & = & 0.006 \max \left( 1 , \frac{2.5}{M} \right) \\
q & = & {\log}_e \left( \frac{R}{R_\rmn{c}} \right) \, .
\end{eqnarray}
The luminosity and radius of the star are then given by $L'$ and $R'$. 

\begin{figure}
\centerline{
\psfig{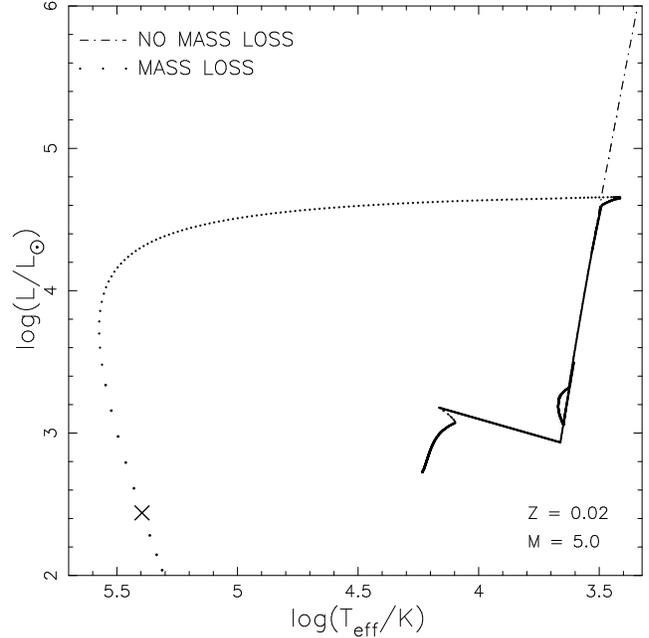}
}
\caption{Synthetic evolution tracks on the HRD for a $5.0 \Msun$ star
without mass loss (dash-dot line) and with mass loss (points). The cross 
marks where the WD cooling track begins.}
\label{f:ffig16}
\end{figure}

\begin{figure}
\centerline{
\psfig{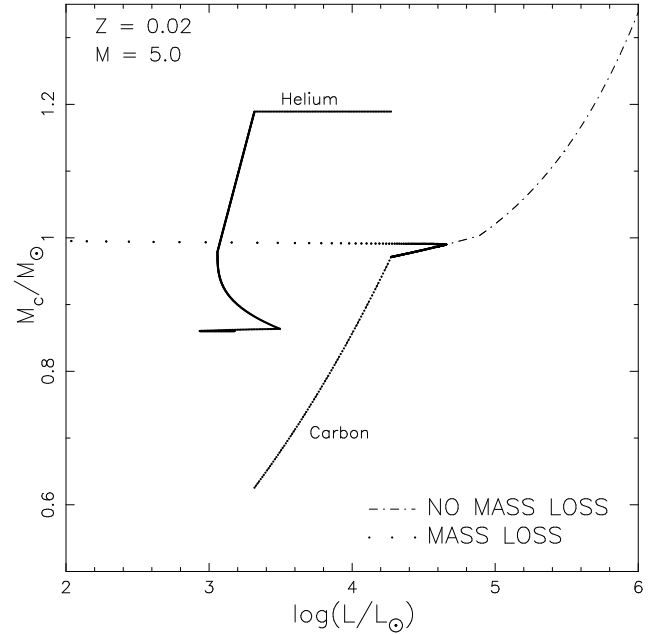}
}
\caption{Relation between core mass and luminosity for a $5.0 \Msun$ star as
given by the formulae without mass loss (dash-dot line) and with mass loss
(points). Both the helium and carbon-oxygen cores are shown for the EAGB
phase.}
\label{f:ffig17}
\end{figure}

In the above formulae, $L_\rmn{c}$ and $R_\rmn{c}$ are the luminosity and 
radius of the remnant that the star would become if it lost all of its 
envelope immediately. 
Thus we set $M = \Mc$ in the appropriate remnant formulae.  
If the star is on the HG or GB then we have, for $M < \MHeF$, 
\begin{eqnarray*}
L_\rmn{c} = \LZHe \left( \Mc \right) \\
R_\rmn{c} = \RZHe \left( \Mc \right) 
\end{eqnarray*}
otherwise
\[ 
L_\rmn{c} = L_\rmn{WD} \left( \Mc \right) , \, \mbox{ i.e. eq.~(\ref{e:Lwd}) 
with $A = 4$ and $t = 0$,} \] 
\[ R_\rmn{c} = R_\rmn{WD} \left( \Mc \right) \, . \] 
During CHeB the remnant will be an evolved helium MS star so we use \Mc\ and 
$\tau = (t - \tHeI)/\tHe$ in eqns.~(\ref{e:LHeMS}) and (\ref{e:RHeMS}) 
to give $L_\rmn{c}$ and $R_\rmn{c}$ respectively.
On the EAGB the remnant will be a helium HG or GB star with $M = \Mche$ so that 
$L_\rmn{c}$ comes from the HeGB \Mc-$L$ relation with $\Mc = \Mcco$ and 
$R_\rmn{c}$ from $R_\rmn{HeGB} = \left( \Mche , L_\rmn{c} \right)$.
For the TPAGB, HeHG and HeGB the remnant will most likely be a CO WD so 
\[ L_\rmn{c} = L_\rmn{WD} \left( \Mc \right) , \, \mbox{ i.e. eq.~(\ref{e:Lwd}) 
with $A = 15$ and $t = 0$,} \] 
\[ R_\rmn{c} = R_\rmn{WD} \left( \Mc \right) \, . \]
Figure~\ref{f:ffig16} shows how a model incorporating mass loss 
(using the prescription outlined in Section~\ref{s:mloss}) and the 
small-envelope 
pertubation functions deviates from a model without either. 
No difference is evident until the stellar wind becomes appreciable as the 
star evolves up the AGB. 
As the envelope mass is reduced the star initially moves to the right of 
the AGB becoming redder in accordance with eq.~(\ref{e:Ragb}). 
Then as the envelope is reduced even further in mass  
the star moves to the left in the HRD, under the influence of the pertubation 
functions, becoming bluer as the hot core starts to become visible.
Thus we have in effect mimicked the planetary nebulae nucleus phase of  
evolution which finishes when the star joins up with the white dwarf cooling 
track (marked by a cross on the figure). 
The behaviour of the core-mass-luminosity relation for the same models is 
shown in Fig.~\ref{f:ffig17}.  
Both the helium and the carbon-oxygen cores are shown on the AGB until 
second dredge-up when the helium core is reduced in mass and the two grow 
together. 
It can be seen that after second dredge-up the slope of the relation 
changes as a result of third dredge-up during the TPAGB phase.

We should note that $R_\rmn{c}$ can be used directly as a fairly 
accurate estimate of the current core radius of the star except when 
$R_\rmn{c}$ is given by $\RWD$.  
In that case nuclear burning will be taking place in a thin shell 
separating the giant core from the envelope so that the core will be 
a hot subdwarf for which we assume the radius 
$R_\rmn{c} \simeq 5 \RWD \left( \Mc \right)$.
It is also necessary to check that $R_\rmn{c} \leq R$ in all cases.

\section{Mass loss and rotation}
\label{s:mlossrt}

\subsection{Mass loss}
\label{s:mloss}

We now describe a particular mass loss prescription which is independent of 
the previous formulae and fits observations well. 
On the GB and beyond, we apply mass loss to the envelope according to
the formula of Kudritzki \& Reimers (1978),
\beq\label{e:mreim}
{\dot{M}}_\rmn{R} = \eta \, 4 \times 10^{-13} \frac{LR}{M} 
\: \Msun {\rmn{yr}}^{-1},
\eeq
with a value of $\eta = 0.5$.
Our value for $\eta$ is within the limits set by observations of Horizontal 
Branch morphology in Galactic globular clusters (Iben \& Renzini 1983) and 
we don't include a $Z$-dependence in eq.~(\ref{e:mreim}) as there is no 
strong evidence that it is necessary (Iben \& Renzini 1983; Carraro et al.\  
1996). 
On the AGB, we apply the formulation of Vassiliadis \& Wood (1993),
\[
\log {\dot{M}}_\rmn{VW} = -11.4 + 0.0125[P_0 - 100 \, \max (M-2.5,0.0)],
\]
to give the observed rapid exponential increase in 
$\dot{M}$ with period 
before the onset of the the superwind phase. 
The steady superwind phase is then modelled by applying a maximum of 
${\dot{M}}_\rmn{VW} = 1.36 \times 10^{-9} L \: \Msun {\rmn{yr}}^{-1}$. 
$P_0$ is the Mira pulsation period given by 
\[
\log P_0 = \min \left( 3.3 , -2.07 - 0.9 \log M + 1.94 \log R 
\right) \, .
\]

For massive stars we model mass loss over the entire HRD using the
prescription given by Nieuwenhuijzen \& de Jager (1990), 
\[ 
{\dot{M}}_\rmn{NJ} = \left( \frac{Z}{Z_{\odot}} \right)^{1/2} 
9.6 \times 10^{-15} R^{0.81} L^{1.24} M^{0.16} \: \Msun {\rmn{yr}}^{-1} 
\]
for $L > 4000 \Lsun$, modified by the factor $Z^{1/2}$ 
(Kudritzki et al.\ 1989). 

For small H-envelope mass, $\mu < 1.0$, we also include a Wolf-Rayet-like mass 
loss (Hamann, Koesterke \& Wessolowski 1995; Hamann \& Koesterke 1998) 
which we have reduced to give 
\[
{\dot{M}}_\rmn{WR} = 10^{-13} L^{1.5} \left( 1.0 - \mu \right) 
\: \Msun {\rmn{yr}}^{-1} 
\]
where $\mu$ is given by eqn.~(\ref{e:muenv}). 
The reduction is necessary in order to produce sufficient black holes to 
match the number observed in binaries. 

We than take the mass loss rate as the dominant mechanism at that time 
\[ \dot{M} = \max \left( {\dot{M}}_\rmn{R} , {\dot{M}}_\rmn{VW} , 
{\dot{M}}_\rmn{NJ} , {\dot{M}}_\rmn{WR} \right) \: \Msun {\rmn{yr}}^{-1} . 
\]

In addition we add a LBV-like mass loss for stars beyond the 
Humphreys-Davidson limit (Humphreys \& Davidson 1994), 
\[
{\dot{M}}_\rmn{LBV} = 0.1 \left( 10^{-5} R L^{1/2} \right)^{3} 
\left( \frac{L}{6 \times 10^5} - 1.0 \right) \: \Msun {\rmn{yr}}^{-1} ,
\]
if $L > 6 \times 10^5$ and $10^{-5} R L^{1/2} > 1.0$, so that 
$\dot{M} = \dot{M} + {\dot{M}}_\rmn{LBV}$. 

For naked helium stars we include the Wolf-Rayet-like mass loss rate to 
give 
\[
\dot{M} = \max \left( {\dot{M}}_\rmn{R} , {\dot{M}}_\rmn{WR} \left( \mu = 0 
\right) \right) 
\: \Msun {\rmn{yr}}^{-1} . 
\]

The introduction of mass loss means that we now have two mass variables, 
the initial mass $M_0$ and the current mass $M_t \left( = M \right)$. 
From tests with mass loss on detailed evolution models we found that the 
luminosity and timescales remain virtually unchanged when mass loss is 
included, during the GB and beyond, but that the radius behaviour is very 
sensitive. 
Thus we use $M_0$ in all formulae that involve the calculation of 
timescales, luminosity or core mass and we use $M_t$ in all radius formulae. 
When a MS star loses mass, which may occur in a stellar wind for massive 
stars or as a result of mass transfer, it will evolve down along the MS to 
lower $L$ and \Teff\ because of the decrease in central density and 
temperature. 
The luminosity responds to changes in mass because the size of the core 
depends on the mass of the star and therefore $M_0$, which is more correctly 
the effective initial mass, is kept equal to the current mass while the 
star is on the MS. 
We must effectively age the star, so that the fraction of MS lifetime 
remains unchanged, by using 
\[
t' = \frac{{\tMS}'}{\tMS} t 
\]
where primes denote quantities after a small amount of mass loss 
(${\tMS}' > \tMS$ thus $t' > t$). 
Even though the star has been aged relative to stars of its new mass, its 
remaining MS lifetime has been increased. 
Naked helium main-sequence stars must also be treated in the same way with
$\tMS$ replaced by $\tHeMS$.
During the giant phases of evolution the age determines the core mass which 
will be unaffected by mass changes at the surface, as the core and envelope 
are effectively decoupled in terms of the stellar structure, so that the 
age and the initial mass do not need to be altered. 
HG stars will respond to changes in mass on a thermal timescale and thus, 
as our detailed models show is necessary, we keep $M_0 = M_t$ during the HG 
and the star is aged according to  
\[
t' = {\tMS}' + \frac{\left( {\tBGB}' - {\tMS}' \right)}{\left( \tBGB - \tMS 
\right)} \left( t - \tMS \right)
\]
whenever mass is lost.
However, as the core mass depends on $M_0$, 
see eqs.~(\ref{e:mc1hg}-\ref{e:mc3hg}), there exists a limiting value beyond 
which $M_0$ cannot be decreased.
To do otherwise would lead to an unphysical decrease in the core mass. 
Therefore our treatment of mass loss on the HG is a mixture of the way the MS 
and giant phases are treated which in a sense reflects the transitional nature
of the HG phase of evolution. 

When a LM star experiences the He-flash and moves to the ZAHB we reset  
$M_0 = M_t$, so that $t = \tHeI \left( M_0 \right)$ as it is now a new star 
with no knowledge of its history. 
We also reset $M_0 = M_t$ when naked helium star evolution is begun.  

\subsubsection{The white dwarf initial--final mass relation}

\begin{figure}
\centerline{
\psfig{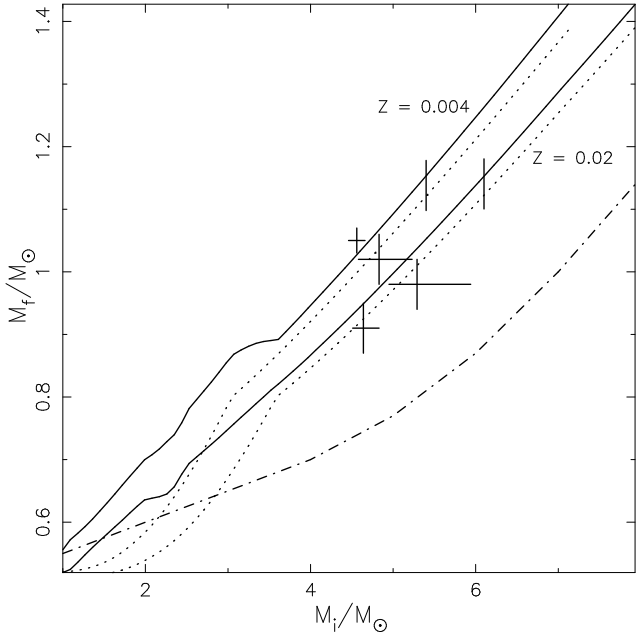}
}
\caption{Relation between white dwarf mass and the ZAMS mass of its' 
progenitor, i.e. the initial${\DS -}$final mass relation (IFMR). 
The IFMR from the evolution formulae (solid line) is given for $Z = 0.02$ 
and $Z = 0.004$ as well as the corresponding core masses 
at the start of the TPAGB (dotted lines). 
The vertical lines correspond to $\Mup$.
Weidemann's (1987) semi-empirical IFMR (dash-dot line) and the NGC~2516 
white dwarfs (crosses) from Jeffries (1997) are shown.} 
\label{f:ffig18}
\end{figure}

If a star is to evolve to become a WD the minimum mass possible for the WD is 
the core mass at the start of the TPAGB. 
Thus an accurate empirical relation between white dwarf masses and the initial 
mass of their progenitors provides an important calibration of the mass 
loss required on the AGB. 
This helps to constrain $\eta$ in eq.~(\ref{e:mreim}) which, for now, is 
basically a free parameter. 
The commonly used method to obtain the initial--final mass relation (IFMR) for 
white dwarfs is to use WDs that are members of clusters with known ages. 
Their radii, masses and cooling times can be obtained spectroscopically so that 
by subtracting the cooling time from the cluster age the time spent by the 
progenitor from the ZAMS to the AGB can be estimated. 
The initial progenitor mass, $M_i$, must then be derived using appropriate 
stellar models so that this a semi-empirical method for defining the IFMR. 
Using data from WDs in galactic open clusters Weidemann (1987) derived such 
a semi-empirical IFMR as shown in Fig.~\ref{f:ffig18}. 
As Jeffries (1997) rightly points out, an IFMR derived by this method will 
be sensitive to the amount of core overshooting included in the stellar 
evolution models. 
The effect of increased overshooting is to decrease the derived cluster age, 
thus increasing the progenitor lifetime and decreasing $M_i$. 
The IFMR will also be sensitive to changes in metallicity. 

Jeffries (1997) presents initial and final masses for 4 WDs found in the young 
open cluster NGC~2516 which has a metallicity of $Z \simeq 0.009$. 
The initial progenitor masses are derived from the stellar models of 
Schaerer et al.\ (1993) with $Z = 0.008$ and moderate core overshooting. 
We show the data points for these 4 WDs in Fig.~\ref{f:ffig18} 
as well as the IFMR given by our formulae for $Z = 0.02$ and $Z = 0.004$ 
(the IFMR for $Z = 0.009$ will lie between these two), 
and the corresponding core mass at the start of the TPAGB. 
As the TPAGB core mass is the minimum possible mass for the WD it is clear 
that our formulae are in disagreement with the semi-empirical IFMR of 
Weidemann (1987). 
Jeffries (1997) was in similar disagreement with the semi-empirical IFMR. 
However the IFMRs from our formulae are in good agreement with the NGC~2516 
data, taking the associated errors of the data points into account. 
Thus there is no contradiction with the mass loss prescription used for the 
formulae however, we note that an empirical IFMR is required before 
concrete conclusions can be drawn.

\subsection{Rotation}
\label{s:rotat}

As we plan to use the evolution routines for single stars in binary star 
applications it is desirable to follow the evolution of the stars' angular 
momentum. 
To do this we must start each star with some realistic spin on the ZAMS. 
A reasonable fit to the ${\bar{v}}_\rmn{rot}$ MS data of Lang (1992) is given 
by 
\beq
{\bar{v}}_\rmn{rot} \left( M \right) = \frac{330 M^{3.3}}{15.0 + M^{3.45}}
 \: \rmn{km} \, {\rmn{s}}^{-1}
\eeq
so that
\beq
\Omega = 45.35 \frac{{\bar{v}}_\rmn{rot}}{\RZAMS} \quad \rmn{yr^{-1}} \, .
\eeq
The angular momentum is then given by 
\[ J_\rmn{spin} = I \Omega = k M R^2 \Omega \] 
where the constant $k$ depends on the internal structure, e.g. $k = 2/5$ for 
a solid sphere and $k = 2/3$ for a spherical shell. 
In actual fact we find the angular momentum by splitting the star 
into two parts, consisting of the core and the envelope, so that 
\beq
J_\rmn{spin} = \left( k_2 \left( M - \Mc \right) R^2 + k_3 \Mc {R_\rmn{c}}^2 
\right) \Omega  
\eeq
where $k_2 = 0.1$, based on detailed giant models which reveal
$k = 0.1 \Menv / M$, and $k_3 = 0.21$ for an $n = 3/2$ polytrope 
such as a WD, NS or dense convective core. 
This works well for post-MS stars which have developed a dense core whose 
rotation is likely to have decoupled from the envelope while also 
representing the near uniform rotation of homogenous MS stars which have 
$\Mc = 0.0$. 
When the star loses mass in a stellar wind the wind will carry off 
angular momentum from the star at a rate given by 
\[ \dot{J} = k \dot{M} h \] 
where $h = R^2 \Omega$. 
Thus 
\beq
J_\rmn{spin} = J_\rmn{spin} - \frac{2}{3} \Delta M R^2 \Omega 
\eeq
when the star loses an amount of mass $\Delta M$, where we take $k = 2/3$ as 
we assume that all the mass is lost uniformly at the surface of the star, 
ie. from a spherical shell. 

We also include magnetic braking for stars that have appreciable 
convective envelopes where 
\beq
{\dot{J}}_\rmn{mb} = 5.83 \times 10^{-16} \frac{\Menv}{M} 
\left( R \Omega \right)^{3} \: \Msun {\rmn{R}}^2_{\odot} {\rmn{yr}}^{-2} , 
\eeq
with $\Omega$ in units of years. 
However, following Rappaport et al.\ (1983), we don't allow magnetic braking  
for fully convective stars, $M < 0.35$.

For most stars \Menv\ is simply given by $M - \Mc$ however the case is slightly 
more complicated for MS and HG stars. 
Our detailed models show that MS stars are fully convective for $M < 0.35$ so 
that $M_{\rmn{env},0} = M$ and that MS stars with $M > 1.25$ have little or no 
convective envelope so that $M_{\rmn{env},0} = 0.0$, independent of $Z$.
In between we take 
\[ M_{\rmn{env},0} = 0.35 \left( \frac{1.25 - M}{0.9} \right)^{2} \quad 
0.35 \leq M \leq 1.25  \, . \]
The convective envelope, if it is present, will diminish as the star 
evolves across the MS so we take  
\[ \Menv = M_{\rmn{env},0} \left( 1.0 - \tau \right)^{1/4} \, , \] 
where 
\[ \tau = \frac{t}{\tMS} \]
and $M_{\rmn{env},0}$ is effectively the ZAMS value. 
On the HG we assume that the convective core gradually establishes itself 
so that 
\[ \Menv = \tau \left( M - \Mc \right) \]
where 
\[ \tau = \frac{t - \tMS}{\tBGB - \tMS} \, . \]

\section{Discussion}
\label{s:disc}

\begin{figure*}
\centerline{
\psfig{figure=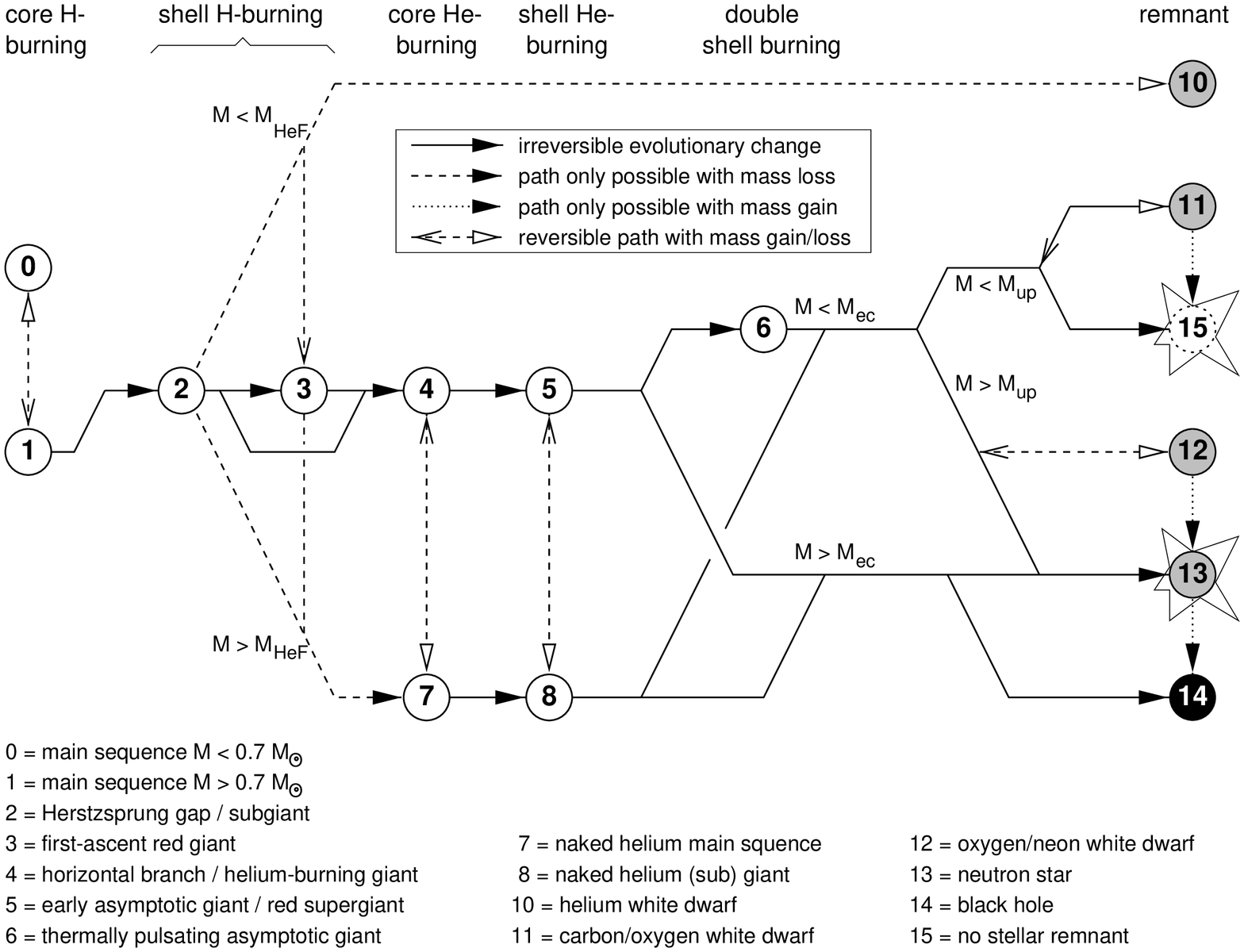,width=160mm,angle=+90}
}
\caption{}
\label{f:ffig19}
\end{figure*}

\begin{figure}
\centerline{
\psfig{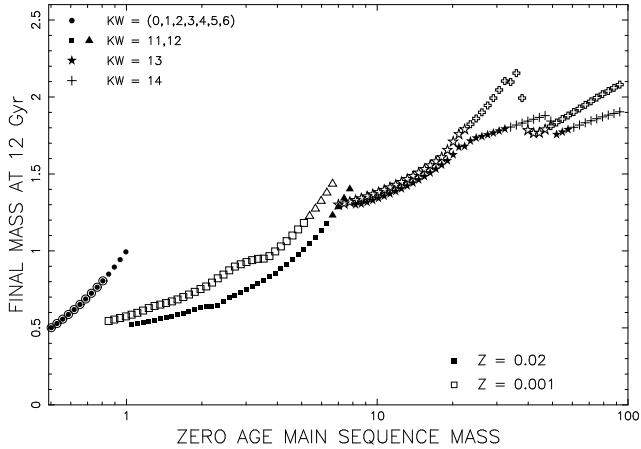}
}
\caption{Distribution of remnant masses and types after $1.2 \times 10^{10}$ 
yrs of evolution, as a function of 
initial mass, for $Z = 0.001$ (hollow symbols) and $Z = 0.02$ 
(filled symbols).}
\label{f:ffig20}
\end{figure}

The possible paths of evolution through the various phases described in the 
preceeding sections are illustrated in Fig.~\ref{f:ffig19}. 
In Fig.~\ref{f:ffig20} we show the distribution of remnant masses and types, 
as a function of initial stellar mass, for Population I and II stars 
as given by the rapid evolution code. 
The distribution approximates what we would see if a population of single 
stars were to be evolved to the current age of the Galaxy. 
The variation in behaviour produced by a change in metallicity should once 
again be noted. 
These variations are due to changes in the evolution rates as a function of 
initial mass, brought about by changes in the composition. 
The initial mass above which stars will become black holes rather than 
neutron stars is not well constrained which is why we use the 
maximum AGB core mass in the formulae to decide the outcome, 
corresponding to a transition at $M_0 \simeq 30 \Msun$ 
(varying with metallicity).
It can also be seen from Fig.~\ref{f:ffig20} that, above this mass, a small 
pocket of neutron star formation occurs in what would normally be assumed to 
be a region of black hole formation on the diagram. 
This behaviour corresponds to a massive star losing its envelope on the HG 
so that the star enters the naked helium MS phase, where the mass 
loss rate increases, causing a reduction in $M_0$. 
As a result a lower value than otherwise expected for $\MNS$ is given by 
eq.~(\ref{e:Mns}) when the naked helium evolution ends. 

The formulae described in this paper are available in convenient 
subroutine form as a SSE package, which we also term `the rapid evolution 
code', that contains:
\begin{description}
\item[EVOLVE] The main routine which, amongst other things, initialises the 
 star, chooses the timesteps and implements mass loss.
\item[ZCNSTS] Subroutine which sets all the constants of the formulae which 
depend on metallicity so that there is no $Z$ dependence elsewhere. This needs 
to be called each time $Z$ is changed.
\item[STAR  ] Subroutine which derives the landmark timescales and 
luminosities that divide the various evolution stages. It also calculates 
$t_\rmn{N}$ which is an estimate of the end of the nuclear evolution, ie. 
when $\Mc = \min \left( M_t , \McSN \right)$, assuming no further mass 
loss.
\item[HRDIAG] Subroutine to decide which evolution stage the star is 
currently at and then to calculate the appropriate $L$, $R$ and \Mc.
\item[ZFUNCS] Contains all the detailed evolution formulae as functions.
\item[MLWIND] derives the mass loss as a function of evolution stage 
and the current stellar properties.
\end{description}
In the absence of mass loss STAR is only required at the 
beginning of the evolution and then HRDIAG can be called at any age to return 
the correct stellar quantities. 
When mass loss is included, HRDIAG must be called often enough that only a 
small amount of mass is lost during each timestep. 
STAR also needs to be called often as some timescales need to be reset 
after changes of type, e.g. start of the HeMS, as do some luminosities, 
e.g. \LZAHB\ depends on the envelope mass at the He-flash. 

The following timesteps, $\delta t_k$, are assigned according to the stellar 
type, $k$:
\[
\delta t_k =  \left\{ \begin{array} {l@{\quad}l}
\frac{1}{100} t_\rmn{MS} & k = 0, 1 \\
\frac{1}{20} \left( t_\rmn{BGB} - t_\rmn{MS} \right) & k = 2 \\
\frac{1}{50} \left( t_\rmn{inf,1} - t \right) & k = 3 \: t \leq t_x \\
\frac{1}{50} \left( t_\rmn{inf,2} - t \right) & k = 3 \: t > t_x \\
\frac{1}{50} t_\rmn{He} & k = 4 \\
\frac{1}{50} \left( t_\rmn{inf,1} - t \right) & k = 5, 6 \: t \leq t_x \\
\frac{1}{50} \left( t_\rmn{inf,2} - t \right) & k = 5, 6 \: t > t_x \\
\frac{1}{20} t_\rmn{HeMS} & k = 7 \\
\frac{1}{50} \left( t_\rmn{inf,1} - t \right) & k = 8, 9 \: t \leq t_x \\
\frac{1}{50} \left( t_\rmn{inf,2} - t \right) & k = 8, 9 \: t > t_x \\
\max \left( 0.1 , 10.0 t \right) & k \geq 10 
\end{array} \right. \, . \]
In addition we impose a maximum TPAGB timestep of $5 \times 10^{-3}$ Myr so 
that important contributions from the small-envelope pertubation functions 
are not missed. 
We also calculate $\delta t_e$, the time to the next change of stellar type 
(e.g. $\delta t_e = t_\rmn{MS} - t$ for $k$ = 0,1), and $\delta t_\rmn{N}$ 
which is the current remaining nuclear lifetime of the star (i.e. 
$\delta t_\rmn{N} = t_\rmn{N} - t$ assuming that the star is in a nuclear 
burning stage, otherwise $t_\rmn{N}$ is set to some large dummy value).  
If necessary we limit the timestep such that mass loss will be less 
than 1\% over the timestep, 
\[ \delta t_\rmn{ml} = -0.01 \frac{M}{\dot{M}} \, , \]
and we also limit the timestep so that the radius will not change by more 
than 10\%, 
\[ \delta t_{R} = 0.1 \frac{R}{| \dot{R} |} \, . \]
Therefore the timestep is given by 
\beq\label{e:tstep}
\delta t = \min \left( \delta t_k , \delta t_e  , \delta t_\rmn{N} , 
\delta t_\rmn{ml} , \delta t_{R} \right) \, . 
\eeq 
In some cases the choice of timesteps is purely for aesthetic purposes so the 
size could easily be increased with no loss of accuracy if extra speed is 
required, such as for evolving large stellar populations. 
For example, the MS can be safely done in one timestep but then, for an 
individual star, the hook 
feature would not appear on a HRD plotted from the resulting output.  

Using the SSE package we can evolve 10000 stars up to the age of the Galaxy 
in approximately $100 \,$s of cpu time on a Sun SparcUltra10 workstation 
(containing a 300 MHz processor). 
Thus a million stars can be evolved in roughly the time taken to compute 
one detailed model track. 
This speed coupled with the accuracy of the formulae make the SSE package 
ideal for any project that requires information derived from the evolution 
of a large number of stars. 
However, the formulae do not render the model grid of Pols et al.~(1998) 
redundant as it contains 
a wealth of information detailing the interior structure of each star, 
information that the formulae simply cannot provide. 
In actual fact the two approaches complement one another. 

The evolution formulae described in this paper have been incorporated into a 
rapid binary evolution algorithm so that we can conduct population synthesis 
involving single stars and binaries. 
The SSE subroutines have also been added to an $N$-body code for the simulation 
of cluster populations. 
In the future we plan to make $\dov$ a free parameter as a variable 
amount of convective overshooting may be preferable, especially in the mass 
range of $1.0$ to $2.0 \Msun$. 
Formulae that describe surface element abundances will also be added so 
that the rapid evolution code can be used for nucleosynthesis 
calculations.

To obtain a copy of the SSE package described in this paper send a request 
to the authors who will provide the {\sc fortran} subroutines by ftp. 

\section*{ACKNOWLEDGMENTS}

JRH thanks Trinity College and the Cambridge Commonwealth Trust for their 
kind support. 
CAT is very grateful to PPARC for support from an advanced fellowship. 
ORP thanks the Institute of Astronomy, Cambridge for supporting a number of 
visits undertaken during this work. 
We would like to thank Peter Eggleton for many helpful suggestions and 
Sverre Aarseth for useful comments.

\newpage
\onecolumn

\section*{APPENDIX}

The $Z$ dependence of the coefficients $a_n$ and $b_n$ is given here.
Unless otherwise stated
\[ a_n = \alpha + \beta \zeta + \gamma {\zeta}^2 + \eta {\zeta}^3 +
\mu {\zeta}^4 \, , \]
and similarly for $b_n$,
where
\[ \zeta = \log(Z/0.02) \, . \]
The variables
\[ \sigma = \log(Z) \]
and
\[ \rho = \zeta + 1.0 \]
are also used.

\vspace{0.5cm}
\begin{tabular}[h]{|lccccc|}
\hline
  & $\alpha$ & $\beta$ & $\gamma$ & $\eta$ & $\mu$ \\
\hline
 $a_{\SSS 1}$ & 1.593890(+3)  & 2.053038(+3)  & 1.231226(+3)  & 2.327785(+2)  
&   \\
 $a_{\SSS 2}$ & 2.706708(+3)  & 1.483131(+3)  & 5.772723(+2)  & 7.411230(+1)  
&   \\
 $a_{\SSS 3}$ & 1.466143(+2)  & -1.048442(+2)  & -6.795374(+1)  & -1.391127(+1)
&   \\
 $a_{\SSS 4}$ & 4.141960(-2)  & 4.564888(-2)  & 2.958542(-2)  & 5.571483(-3)  
&   \\
 $a_{\SSS 5}$ & 3.426349(-1)  &   &   &   &   \\
\hline
 $a_{\SSS 6}$ & 1.949814(+1)  & 1.758178(+0)  & -6.008212(+0)  & -4.470533(+0) 
&  \\
 $a_{\SSS 7}$ & 4.903830(+0)  &   &   &   &   \\
 $a_{\SSS 8}$ & 5.212154(-2)  & 3.166411(-2)  & -2.750074(-3)  & -2.271549(-3) 
&  \\
 $a_{\SSS 9}$ & 1.312179(+0)  & -3.294936(-1)  & 9.231860(-2)  & 2.610989(-2) 
&   \\
 $a_{\SSS 10}$ & 8.073972(-1)  &   &   &   &   \\
\hline
 $a_{\SSS 11}'$ & 1.031538(+0)  & -2.434480(-1)  & 7.732821(+0)  & 6.460705(+0) 
& 1.374484(+0)  \\
 $a_{\SSS 12}'$ & 1.043715(+0)  & -1.577474(+0)  & -5.168234(+0)  & 
-5.596506(+0)  & -1.299394(+0)  \\ 
 $a_{\SSS 13}$ & 7.859573(+2)  & -8.542048(+0)  & -2.642511(+1)  & 
-9.585707(+0)  & \\ 
 $a_{\SSS 14}$ & 3.858911(+3)  & 2.459681(+3)  & -7.630093(+1)  & 
-3.486057(+2)  & -4.861703(+1)  \\ 
 $a_{\SSS 15}$ & 2.888720(+2)  & 2.952979(+2)  & 1.850341(+2)  & 3.797254(+1) 
&   \\
 $a_{\SSS 16}$ & 7.196580(+0)  & 5.613746(-1)  & 3.805871(-1)  & 8.398728(-2) 
&   \\
\hline
\end{tabular}
\[ \begin{array} {l}
a_{\SSS 11} = a_{\SSS 11}' a_{\SSS 14} \\
a_{\SSS 12} = a_{\SSS 12}' a_{\SSS 14}
\end{array} \]

\vspace{0.5cm}
\begin{tabular}[h]{|lccccc|}
\hline
  & $\alpha$ & $\beta$ & $\gamma$ & $\eta$ & $\mu$ \\
\hline
 $a_{\SSS 18}'$ & 2.187715(-1) & -2.154437(+0) & -3.768678(+0) & -1.975518(+0) &
-3.021475(-1) \\
 $a_{\SSS 19}'$ & 1.466440(+0) & 1.839725(+0) & 6.442199(+0) & 4.023635(+0) &
6.957529(-1) \\
 $a_{\SSS 20}$ & 2.652091(+1) & 8.178458(+1) & 1.156058(+2) & 7.633811(+1) &
1.950698(+1) \\
 $a_{\SSS 21}$ & 1.472103(+0) & -2.947609(+0) & -3.312828(+0) & -9.945065(-1) 
&   \\
 $a_{\SSS 22}$ & 3.071048(+0) & -5.679941(+0) & -9.745523(+0) & -3.594543(+0) 
&  \\
 $a_{\SSS 23}$ & 2.617890(+0) & 1.019135(+0) & -3.292551(-2) & -7.445123(-2) 
&  \\
 $a_{\SSS 24}$ & 1.075567(-2) & 1.773287(-2) & 9.610479(-3) & 1.732469(-3) & \\
 $a_{\SSS 25}$ & 1.476246(+0) & 1.899331(+0) & 1.195010(+0) & 3.035051(-1) & \\
 $a_{\SSS 26}$ & 5.502535(+0) & -6.601663(-2) & 9.968707(-2) & 3.599801(-2) & \\
\hline
\end{tabular}
\[ \begin{array} {l}
\log a_{\SSS 17} = \max \left( 0.097 - 0.1072 \left( \sigma + 3 \right)
, \max \left( 0.097 , \min \left( 0.1461 , 0.1461 + 0.1237 \left(
\sigma + 2 \right) \right) \right) \right) \\
a_{\SSS 18} = a_{\SSS 18}' a_{\SSS 20} \\
a_{\SSS 19} = a_{\SSS 19}' a_{\SSS 20}
\end{array} \]

\vspace{0.5cm}
\begin{tabular}[h]{|lcccc|}
\hline
  & $\alpha$ & $\beta$ & $\gamma$ & $\eta$ \\
\hline
 $a_{\SSS 27}$ & 9.511033(+1) & 6.819618(+1) & -1.045625(+1) & -1.474939(+1) \\
 $a_{\SSS 28}$ & 3.113458(+1) & 1.012033(+1) & -4.650511(+0) & -2.463185(+0) \\
 $a_{\SSS 29}'$ & 1.413057(+0) & 4.578814(-1) & -6.850581(-2) & -5.588658(-2) \\
 $a_{\SSS 30}$ & 3.910862(+1) & 5.196646(+1) & 2.264970(+1) & 2.873680(+0) \\
 $a_{\SSS 31}$ & 4.597479(+0) & -2.855179(-1) & 2.709724(-1) & \\
 $a_{\SSS 32}$ & 6.682518(+0) & 2.827718(-1) & -7.294429(-2) & \\
\hline
\end{tabular}
\[ \begin{array} {l}
a_{\SSS 29} = {a_{\SSS 29}'}^{{\TS a}_{\SSS 32}}
\end{array} \]

\vspace{0.5cm}
\begin{tabular}[h]{|lcccc|}
\hline
  & $\alpha$ & $\beta$ & $\gamma$ & $\eta$ \\
\hline
 $a_{\SSS 34}$ & 1.910302(-1) & 1.158624(-1) & 3.348990(-2) & 2.599706(-3) \\
 $a_{\SSS 35}$ & 3.931056(-1) & 7.277637(-2) & -1.366593(-1) & -4.508946(-2) \\
 $a_{\SSS 36}$ & 3.267776(-1) & 1.204424(-1) & 9.988332(-2) & 2.455361(-2) \\
 $a_{\SSS 37}$ & 5.990212(-1) & 5.570264(-2) & 6.207626(-2) & 1.777283(-2) \\
\hline
\end{tabular}
\[ \begin{array} {l}
a_{\SSS 33} = \min \left( 1.4 , 1.5135 + 0.3769 \zeta \right) \\
a_{\SSS 33} = \max \left( 0.6355 - 0.4192 \zeta , \max
\left( 1.25 , a_{\SSS 33} \right) \right)
\end{array} \]

\vspace{0.5cm}
\begin{tabular}[h]{|lcccc|}
\hline
  & $\alpha$ & $\beta$ & $\gamma$ & $\eta$ \\
\hline
 $a_{\SSS 38}$ & 7.330122(-1) & 5.192827(-1) & 2.316416(-1) & 8.346941(-3) \\
 $a_{\SSS 39}$ & 1.172768(+0) & -1.209262(-1) & -1.193023(-1) & -2.859837(-2) \\
 $a_{\SSS 40}$ & 3.982622(-1) & -2.296279(-1) & -2.262539(-1) & -5.219837(-2) \\
 $a_{\SSS 41}$ & 3.571038(+0) & -2.223625(-2) & -2.611794(-2) & -6.359648(-3) \\
 $a_{\SSS 42}$ & 1.9848(+0) & 1.1386(+0) & 3.5640(-1) & \\
 $a_{\SSS 43}$ & 6.300(-2) & 4.810(-2) & 9.840(-3) & \\
 $a_{\SSS 44}$ & 1.200(+0) & 2.450(+0) &  & \\
\hline
\end{tabular}
\[ \begin{array} {l}
a_{\SSS 42} = \min \left( 1.25 , \max \left( 1.1 , a_{\SSS 42} \right)
\right) \\
a_{\SSS 44} = \min \left( 1.3 , \max \left( 0.45 , a_{\SSS 44} \right)
\right)
\end{array} \]

\vspace{0.5cm}
\begin{tabular}[h]{|lcccc|}
\hline
  & $\alpha$ & $\beta$ & $\gamma$ & $\eta$ \\
\hline
 $a_{\SSS 45}$ & 2.321400(-1) & 1.828075(-3) & -2.232007(-2) & -3.378734(-3) \\
 $a_{\SSS 46}$ & 1.163659(-2) & 3.427682(-3) & 1.421393(-3) & -3.710666(-3) \\
 $a_{\SSS 47}$ & 1.048020(-2) & -1.231921(-2) & -1.686860(-2) & -4.234354(-3) \\
 $a_{\SSS 48}$ & 1.555590(+0) & -3.223927(-1) & -5.197429(-1) & -1.066441(-1) \\
 $a_{\SSS 49}$ & 9.7700(-2) & -2.3100(-1) & -7.5300(-2) & \\
 $a_{\SSS 50}$ & 2.4000(-1) & 1.8000(-1) & 5.9500(-1) &  \\
 $a_{\SSS 51}$ & 3.3000(-1) & 1.3200(-1) & 2.1800(-1) &  \\
 $a_{\SSS 52}$ & 1.1064(+0) & 4.1500(-1) & 1.8000(-1) &  \\
 $a_{\SSS 53}$ & 1.1900(+0) & 3.7700(-1) & 1.7600(-1) &  \\
\hline
\end{tabular}
\[ \begin{array} {l}
a_{\SSS 49} = \max \left( a_{\SSS 49} , 0.145 \right) \\
a_{\SSS 50} = \min \left( a_{\SSS 50} , 0.306 + 0.053 \zeta \right) \\
a_{\SSS 51} = \min \left( a_{\SSS 51} , 0.3625 + 0.062 \zeta \right) \\
a_{\SSS 52} = \max \left( a_{\SSS 52} , 0.9 \right) \\
a_{\SSS 52} = \min \left( a_{\SSS 52} , 1.0 \right) \quad \rmn{for} \: Z>0.01 \\
a_{\SSS 53} = \max \left( a_{\SSS 53} , 1.0 \right) \\
a_{\SSS 53} = \min \left( a_{\SSS 53} , 1.1 \right) \quad \rmn{for} \: Z > 0.01
\end{array} \]

\vspace{0.5cm}
\begin{tabular}[h]{|lccccc|}
\hline
  & $\alpha$ & $\beta$ & $\gamma$ & $\eta$ & $\mu$ \\
\hline
 $a_{\SSS 54}$ & 3.855707(-1) & -6.104166(-1) & 5.676742(+0) & 1.060894(+1) &
5.284014(+0) \\
 $a_{\SSS 55}$ & 3.579064(-1) & -6.442936(-1) & 5.494644(+0) & 1.054952(+1) &
5.280991(+0) \\
 $a_{\SSS 56}$ & 9.587587(-1) & 8.777464(-1) & 2.017321(-1) &  &  \\
 $a_{\SSS 57}$ & 1.5135(+0) & 3.7690(-1) &  &  &  \\
\hline
\end{tabular}
\[ \begin{array} {l}
a_{\SSS 57} = \min \left( 1.4 , a_{\SSS 57} \right) \\
a_{\SSS 57} = \max \left( 0.6355 - 0.4192 \zeta , \max \left( 1.25 ,
a_{\SSS 57} \right) \right)
\end{array} \]

\vspace{0.5cm}
\begin{tabular}[h]{|lccccc|}
\hline
  & $\alpha$ & $\beta$ & $\gamma$ & $\eta$ & $\mu$ \\
\hline
 $a_{\SSS 58}$ & 4.907546(-1) & -1.683928(-1) & -3.108742(-1) & -7.202918(-2) 
&  \\
 $a_{\SSS 59}$ & 4.537070(+0) & -4.465455(+0) & -1.612690(+0) & -1.623246(+0) 
&  \\
 $a_{\SSS 60}$ & 1.796220(+0) & 2.814020(-1) & 1.423325(+0) & 3.421036(-1) &  \\
 $a_{\SSS 61}$ & 2.256216(+0) & 3.773400(-1) & 1.537867(+0) & 4.396373(-1) &  \\
 $a_{\SSS 62}$ & 8.4300(-2) & -4.7500(-2) & -3.5200(-2) &  &  \\
 $a_{\SSS 63}$ & 7.3600(-2) & 7.4900(-2) & 4.4260(-2) &  &  \\
 $a_{\SSS 64}$ & 1.3600(-1) & 3.5200(-2) &  &  &  \\
 $a_{\SSS 65}$ & 1.564231(-3) & 1.653042(-3) & -4.439786(-3) & -4.951011(-3) &
-1.216530d-03 \\
 $a_{\SSS 66}$ & 1.4770(+0) & 2.9600(-1) &  &  &  \\
 $a_{\SSS 67}$ & 5.210157(+0) & -4.143695(+0) & -2.120870(+0) &  &  \\
 $a_{\SSS 68}$ & 1.1160(+0) & 1.6600(-1) &  &  &  \\
\hline
\end{tabular}
\[ \begin{array} {l}
a_{\SSS 62} = \max \left( 0.065 , a_{\SSS 62} \right) \\
a_{\SSS 63} = \min \left( 0.055 , a_{\SSS 63} \right) \quad \rmn{for} \: 
Z < 0.004 \\
a_{\SSS 64} = \max \left( 0.091 , \min \left( 0.121 , a_{\SSS 64} \right)
\right) \\
a_{\SSS 66} = \max \left( a_{\SSS 66} , \min \left( 1.6 , -0.308 - 1.046 \zeta
\right) \right) \\
a_{\SSS 66} = \max \left( 0.8 , \min \left( 0.8 - 2.0 \zeta , a_{\SSS 66}
\right) \right) \\
a_{\SSS 68} = \max \left( 0.9 , \min \left( a_{\SSS 68} , 1.0 \right) \right) \\
a_{\SSS 64} = B = {\alpha}_{\SSS R} \left( M = a_{\SSS 66} \right)
\quad \rmn{for} \: a_{\SSS 68} > a_{\SSS 66} \\
a_{\SSS 68} = \min \left( a_{\SSS 68} , a_{\SSS 66} \right)
\end{array} \]

\vspace{0.5cm}
\begin{tabular}[h]{|lcccc|}
\hline
  & $\alpha$ & $\beta$ & $\gamma$ & $\eta$ \\
\hline
 $a_{\SSS 69}$ & 1.071489(+0) & -1.164852(-1) & -8.623831(-2) & -1.582349(-2) \\
 $a_{\SSS 70}$ & 7.108492(-1) & 7.935927(-1) & 3.926983(-1) & 3.622146(-2) \\
 $a_{\SSS 71}$ & 3.478514(+0) & -2.585474(-2) & -1.512955(-2) & -2.833691(-3) \\
 $a_{\SSS 72}$ & 9.132108(-1) & -1.653695(-1) &  & 3.636784(-2) \\
 $a_{\SSS 73}$ & 3.969331(-3) & 4.539076(-3) & 1.720906(-3) & 1.897857(-4) \\
 $a_{\SSS 74}$ & 1.600(+0) & 7.640(-1) & 3.322(-1) & \\
\hline
\end{tabular}
\[ \begin{array} {l}
a_{\SSS 72} = \max \left( a_{\SSS 72} , 0.95 \right)\quad \rmn{for} \: Z>0.01 \\
a_{\SSS 74} = \max \left( 1.4 , \min \left( a_{\SSS 74} , 1.6 \right) \right)
\end{array} \]

\vspace{0.5cm}
\begin{tabular}[h]{|lcccc|}
\hline
  & $\alpha$ & $\beta$ & $\gamma$ & $\eta$ \\
\hline
 $a_{\SSS 75}$ & 8.109(-1) & -6.282(-1) &  &  \\
 $a_{\SSS 76}$ & 1.192334(-2) & 1.083057(-2) & 1.230969(+0) & 1.551656(+0) \\
 $a_{\SSS 77}$ & -1.668868(-1) & 5.818123(-1) & -1.105027(+1) & -1.668070(+1) \\
 $a_{\SSS 78}$ & 7.615495(-1) & 1.068243(-1) & -2.011333(-1) & -9.371415(-2) \\
 $a_{\SSS 79}$ & 9.409838(+0) & 1.522928(+0) &  &  \\
 $a_{\SSS 80}$ & -2.7110(-1) & -5.7560(-1) & -8.3800(-2) &  \\
 $a_{\SSS 81}$ & 2.4930(+0) & 1.1475(+0) &  &  \\
\hline
\end{tabular}
\[ \begin{array} {l}
a_{\SSS 75} = \max \left( 1.0 , \min \left( a_{\SSS 75} , 1.27 \right)
\right) \\
a_{\SSS 75} = \max \left( a_{\SSS 75} , 0.6355 - 0.4192 \zeta \right) \\
a_{\SSS 76} = \max  \left( a_{\SSS 76} , -0.1015564 - 0.2161264 \zeta - 
0.05182516 {\zeta}^2 \right) \\
a_{\SSS 77} = \max \left( -0.3868776 -0.5457078 \zeta - 0.1463472 {\zeta}^2 ,
\min \left( 0.0 , a_{\SSS 77} \right) \right) \\
a_{\SSS 78} = \max \left( 0.0 , \min \left( a_{\SSS 78} , 7.454 + 9.046
\zeta \right) \right) \\
a_{\SSS 79} = \min \left( a_{\SSS 79} , \max \left( 2.0 , -13.3 - 18.6
\zeta \right) \right) \\
a_{\SSS 80} = \max \left( 0.0585542 , a_{\SSS 80} \right) \\
a_{\SSS 81} = \min \left( 1.5 , \max \left( 0.4 , a_{\SSS 81} \right) \right)
\end{array} \]

\vspace{0.5cm}
\begin{tabular}[h]{|lccccc|}
\hline
  & $\alpha$ & $\beta$ & $\gamma$ & $\eta$ & $\mu$ \\
\hline
 $b_{1}$ & 3.9700(-1) & 2.8826(-1) & 5.2930(-1) &  &  \\
 $b_{4}$ & 9.960283(-1) & 8.164393(-1) & 2.383830(+0) & 2.223436(+0) &
8.638115(-1) \\
 $b_{5}$ & 2.561062(-1) & 7.072646(-2) & -5.444596(-2) & -5.798167(-2) &
-1.349129(-2) \\
 $b_{6}$ & 1.157338(+0) & 1.467883(+0) & 4.299661(+0) & 3.130500(+0) &
6.992080(-1) \\
 $b_{7}$ & 4.022765(-1) & 3.050010(-1) & 9.962137(-1) & 7.914079(-1) &
1.728098(-1) \\
\hline
\end{tabular}
\[ \begin{array} {l}
b_1 = \min \left( 0.54 , b_1 \right) \\
b_2 = 10^{-4.6739 - 0.9394 \sigma} \\
b_2 = \min \left( \max \left( b_2 , -0.04167 + 55.67 Z \right) ,
0.4771 - 9329.21 Z^{2.94} \right) \\
b_3' = \max \left( -0.1451 , -2.2794 - 1.5175 \sigma - 0.254 {\sigma}^2
\right) \\
b_3 = 10^{{\TS b}_3'} \\
b_3 = \max \left( b_3 , 0.7307 + 14265.1 Z^{3.395} \right) \quad \rmn{for}
\: Z > 0.004 \\
b_4 = b_4 + 0.1231572 {\zeta}^5 \\
b_6 = b_6 + 0.01640687 {\zeta}^5
\end{array} \]

\vspace{0.5cm}
\begin{tabular}[h]{|lccc|}
\hline
  & $\alpha$ & $\beta$ & $\gamma$ \\
\hline
 $b_{9}$ & 2.751631(+3) & 3.557098(+2) & \\
 $b_{10}$ & -3.820831(-2) & 5.872664(-2) & \\
 $b_{11}'$ & 1.071738(+2) & -8.970339(+1) & -3.949739(+1) \\
 $b_{12}$ & 7.348793(+2) & -1.531020(+2) & -3.793700(+1) \\
 $b_{13}'$ & 9.219293(+0) & -2.005865(+0) & -5.561309(-1) \\
\hline
\end{tabular}
\[ \begin{array} {l}
b_{11} = {b_{11}'}^2 \\
b_{13} = {b_{13}'}^2
\end{array} \]

\vspace{0.5cm}
\begin{tabular}[h]{|lccc|}
\hline
  & $\alpha$ & $\beta$ & $\gamma$ \\
\hline
 $b_{14}'$ & 2.917412(+0) & 1.575290(+0) & 5.751814(-1) \\
 $b_{15}$ & 3.629118(+0) & -9.112722(-1) & 1.042291(+0) \\
 $b_{16}'$ & 4.916389(+0) & 2.862149(+0) & 7.844850(-1) \\
\hline
\end{tabular}
\[ \begin{array} {l}
b_{14} = {b_{14}'}^{{\TS b}_{15}} \\
b_{16} = {b_{16}'}^{{\TS b}_{15}} \\
b_{17} = 1.0 \\
b_{17} = 1.0 - 0.3880523 \left( \zeta + 1.0 \right)^{2.862149} \quad \rmn{for}
\: \zeta > -1.0
\end{array} \]

\vspace{0.5cm}
\begin{tabular}[h]{|lcccc|}
\hline
  & $\alpha$ & $\beta$ & $\gamma$ & $\eta$ \\
\hline
 $b_{18}$ & 5.496045(+1) & -1.289968(+1) & 6.385758(+0) &  \\
 $b_{19}$ & 1.832694(+0) & -5.766608(-2) & 5.696128(-2) &  \\
 $b_{20}$ & 1.211104(+2) &  &  &  \\
\hline
 $b_{21}$ & 2.214088(+2) & 2.187113(+2) & 1.170177(+1) & -2.635340(+1) \\
 $b_{22}$ & 2.063983(+0) & 7.363827(-1) & 2.654323(-1) & -6.140719(-2) \\
 $b_{23}$ & 2.003160(+0) & 9.388871(-1) & 9.656450(-1) & 2.362266(-1) \\
\hline
 $b_{24}'$ & 1.609901(+1) & 7.391573(+0) & 2.277010(+1) & 8.334227(+0) \\
 $b_{25}$ & 1.747500(-1) & 6.271202(-2) & -2.324229(-2) & -1.844559(-2) \\
 $b_{27}'$ & 2.752869(+0) & 2.729201(-2) & 4.996927(-1) & 2.496551(-1) \\
 $b_{28}$ & 3.518506(+0) & 1.112440(+0) & -4.556216(-1) & -2.179426(-1) \\
\hline
\end{tabular}
\[ \begin{array} {l}
b_{24} = {b_{24}'}^{{\TS b}_{28}} \\
b_{26} = 5.0 - 0.09138012 Z^{-0.3671407} \\
b_{27} = {b_{27}'}^{2 {\TS b}_{28}}
\end{array} \]

\vspace{0.5cm}
\begin{tabular}[h]{|lcccc|}
\hline
  & $\alpha$ & $\beta$ & $\gamma$ & $\eta$ \\
\hline
 $b_{29}$ & 1.626062(+2) & -1.168838(+1) & -5.498343(+0) &  \\
 $b_{30}$ & 3.336833(-1) & -1.458043(-1) & -2.011751(-2) &  \\
 $b_{31}'$ & 7.425137(+1) & 1.790236(+1) & 3.033910(+1) & 1.018259(+1) \\
 $b_{32}$ & 9.268325(+2) & -9.739859(+1) & -7.702152(+1) & -3.158268(+1) \\
 $b_{33}$ & 2.474401(+0) & 3.892972(-1) &  &  \\
 $b_{34}'$ & 1.127018(+1) & 1.622158(+0) & -1.443664(+0) & -9.474699(-1) \\
\hline
\end{tabular}
\[ \begin{array} {l}
b_{31} = {b_{31}'}^{{\TS b}_{33}} \\
b_{34} = {b_{34}'}^{{\TS b}_{33}}
\end{array} \]

\vspace{0.5cm}
\begin{tabular}[h]{|lcccc|}
\hline
  & $\alpha$ & $\beta$ & $\gamma$ & $\eta$ \\
\hline
 $b_{36}'$ & 1.445216(-1) & -6.180219(-2) & 3.093878(-2) & 1.567090(-2) \\
 $b_{37}'$ & 1.304129(+0) & 1.395919(-1) & 4.142455(-3) & -9.732503(-3) \\
 $b_{38}'$ & 5.114149(-1) & -1.160850(-2) &  &  \\
\hline
\end{tabular}
\[ \begin{array} {l}
b_{36} = {b_{36}'}^4 \\
b_{37} = 4.0 b_{37}' \\
b_{38} = {b_{38}'}^4
\end{array} \]

\vspace{0.5cm}
\begin{tabular}[h]{|lcccc|}
\hline
  & $\alpha$ & $\beta$ & $\gamma$ & $\eta$ \\
\hline
 $b_{39}$ & 1.314955(+2) & 2.009258(+1) & -5.143082(-1) & -1.379140(+0) \\
 $b_{40}$ & 1.823973(+1) & -3.074559(+0) & -4.307878(+0) &  \\
 $b_{41}'$ & 2.327037(+0) & 2.403445(+0) & 1.208407(+0) & 2.087263(-1) \\
 $b_{42}$ & 1.997378(+0) & -8.126205(-1) &  &  \\
 $b_{43}$ & 1.079113(-1) & 1.762409(-2) & 1.096601(-2) & 3.058818(-3) \\
 $b_{44}'$ & 2.327409(+0) & 6.901582(-1) & -2.158431(-1) & -1.084117(-1) \\
\hline
\end{tabular}
\[ \begin{array} {l}
b_{40} = \max \left( b_{40} , 1.0 \right) \\
b_{41} = {b_{41}'}^{{\TS b}_{42}} \\
b_{44} = {b_{44}'}^5
\end{array} \]

\vspace{0.5cm}
\begin{tabular}[h]{|lcccc|}
\hline
  & $\alpha$ & $\beta$ & $\gamma$ & $\eta$ \\
\hline
 $b_{46}$ & 2.214315(+0) & -1.975747(+0) &  &  \\
 $b_{48}$ & 5.072525(+0) & 1.146189(+1) & 6.961724(+0) & 1.316965(+0) \\
 $b_{49}$ & 5.139740(+0) &  &  &  \\
\hline
\end{tabular}
\[ \begin{array} {l}
b_{45} = 1.0 - \left( 2.47162 \rho - 5.401682 {\rho}^2 + 3.247361 {\rho}^3
\right) \\
b_{45} = 1.0 \quad \rmn{for} \: \rho \leq 0.0 \\
b_{46} = -1.0 b_{46} \log \left( \frac{\MHeF}{\MFGB} \right) \\
b_{47} = 1.127733 \rho + 0.2344416 {\rho}^2 - 0.3793726 {\rho}^3
\end{array} \]

\vspace{0.5cm}
\begin{tabular}[h]{|lccccc|}
\hline
  & $\alpha$ & $\beta$ & $\gamma$ & $\eta$ & $\mu$ \\
\hline
 $b_{51}'$ & 1.125124(+0) & 1.306486(+0) & 3.622359(+0) & 2.601976(+0) &
3.031270(-1) \\
 $b_{52}$ & 3.349489(-1) & 4.531269(-3) & 1.131793(-1) & 2.300156(-1) &
7.632745(-2) \\
 $b_{53}'$ & 1.467794(+0) & 2.798142(+0) & 9.455580(+0) & 8.963904(+0) &
3.339719(+0) \\
 $b_{54}$ & 4.658512(-1) & 2.597451(-1) & 9.048179(-1) & 7.394505(-1) &
1.607092(-1) \\
 $b_{55}$ & 1.0422(+0) & 1.3156(-1) & 4.5000(-2) &  &  \\
 $b_{56}'$ & 1.110866(+0) & 9.623856(-1) & 2.735487(+0) & 2.445602(+0) &
8.826352(-1) \\
 $b_{57}'$ & -1.584333(-1) & -1.728865(-1) & -4.461431(-1) & -3.925259(-1) &
-1.276203(-1) \\
\hline
\end{tabular}
\[ \begin{array} {l}
b_{51} = b_{51}' - 0.1343798 {\zeta}^5 \\
b_{53} = b_{53}' + 0.4426929 {\zeta}^5 \\
b_{55} = \min \left( 0.99164 - 743.123 Z^{2.83} , b_{55} \right) \\
b_{56} = b_{56}' + 0.1140142 {\zeta}^5 \\
b_{57} = b_{57}' - 0.01308728 {\zeta}^5
\end{array} \]

Note that x($n$) for some number x represents $x \times 10^n$.

A blank entry in a table implies a zero value. 

\bigskip

\bsp 

\label{lastpage}

\begin{thebibliography}{}
\bibitem{} Aarseth S.J., 1996, in Hut P., Makino J., eds, Proc. IAU Symp.
  174, Dynamical Evolution of Star Clusters. Kluwer, Dordrecht, p.  161
\bibitem{} Baglin A., 1997, 23rd meeting of the IAU, Joint Discussion 14, 
Kyoto, Japan
\bibitem{} de Boer K.S., Tucholke H.-J., Schmidt J.H.K., 1997, A\&A, 317, L23
\bibitem{} Carr B.J., Hawking S.W., 1974, Nature, 248, 30 
\bibitem{} Carraro G., Girardi L., Bressan A., Chiosi C., 1996, A\&A, 305, 849 
\bibitem{} Charbonnel C., Meynet G., Maeder A., Schaller G., Schaerer D., 
1993, A\&AS, 101, 415
\bibitem{} Eggleton P.P., 1996, in Hut P., Makino J., eds, Proc. IAU Symp.
             174, Dynamical Evolution of Star Clusters. Kluwer, Dordrecht, p.
             213
\bibitem{} Eggleton P.P., Fitchett M., Tout C.A., 1989, ApJ, 347, 998
\bibitem{} Elson R.A.W., Sigurdsson S., Hurley J.R., Davies M.B., 
Gilmore G.F., 1998, ApJ, 499, L53
\bibitem{} Frost C.A., 1997, PhD Thesis, Monash University
\bibitem{} Hamann W.-R., Koesterke L., 1998, A\&A, 335, 1003
\bibitem{} Hamann W.-R., Koesterke L., Wessolowski U., 1995, A\&A, 299, 151
\bibitem{} Hoffleit D., 1983, The Bright Star Catalogue 4th ed., Yale 
University Observatory, New Haven
\bibitem{} Humphreys R.M., Davidson K., 1994, PASP, 106, 1025
\bibitem{} Iben I.Jr., Renzini A., 1983, ARA\&A, 21, 271 
\bibitem{} Jeffery C.S., 1988, MNRAS 235, 1287
\bibitem{} Jeffries R.D., 1997, MNRAS, 288, 585
\bibitem{} Karakas A., Tout C.A., Lattanzio J.C., 1999, MNRAS, in prep.
\bibitem{} Kudritzki R.P., Pauldrach A., Puls J., Abbott D.C., 1989, A\&A, 
219, 205 
\bibitem{} Kudritzki R.P., Reimers D., 1978, A\&A, 70, 227 
\bibitem{} Lang K.R., 1992, Astrophysical Data, Springer-Verlag
\bibitem{} Lattanzio J.C., 1989, ApJ, 344, L25
\bibitem{} Mestel L., 1952, MNRAS, 112, 583
\bibitem{} Mowlavi N., Schaerer D., Meynet G., Bernasconi P.A., 
Charbonnel C., Maeder A., 1998, A\&AS, 128, 471
\bibitem{} Nieuwenhuizen H., de Jager C., 1990, A\&A, 231, 134 
\bibitem{} Noels A., Fraipont-Caro D., Gabriel M., Grevesse N., Demarque P., 
eds., 1995, Proc. 32nd Li\`{e}ge Int. Astrophys. Colloq., 
Universit\'{e} de Li\`{e}ge 
\bibitem{} Nomoto K., 1984, ApJ, 277, 791
\bibitem{} Perryman M.A.C., H{\o}g E., Kovalovsky J., Lindegren L., Turon C., 
1997, ESA SP-1200
\bibitem{} Pols O.R., Tout C.A., Schr\"oder K.-P., Eggleton P.P., Manners
  J., 1997, MNRAS, 289, 869
\bibitem{} Pols O.R., Schr\"oder K.P., Hurley J.R., Tout C.A., Eggleton 
  P.P., 1998, MNRAS, 298, 525
\bibitem{} Rappaport., Verbunt., Joss., 1983, ApJ, 275, 713
\bibitem{} Schaerer D., Meynet G., Maeder A., Schaller G., 1993, 
             A\&AS, 98, 523
\bibitem{} Schaller G., Schaerer D., Meynet G., Maeder A., 1992,
             A\&AS, 96, 269
\bibitem{} Schr\"oder K.P., Pols O.R., Eggleton P.P., 1997, MNRAS, 285, 696 
\bibitem{} Shapiro S.L., Teukolsky S.A., 1983, Black Holes, White Dwarfs 
             and Neutron Stars, Wiley
\bibitem{} Tout C.A., Pols O.R., Eggleton P.P., Han Z., 1996, MNRAS, 281,
  257 
\bibitem{} Tout C.A., Aarseth S.J., Pols O.R., Eggleton P.P., 1997, MNRAS,
  291, 732
\bibitem{} Van Eck S., Jorissen A., Udry S., Mayor M., Perrier B., 1998, 
A\&A, 329, 971
\bibitem{} Vassiliadis E., Wood P.R., 1993, ApJ, 413, 641
\bibitem{} Weidemann V., 1987, A\&A, 188, 74
\end{thebibliography}
\end{document}